\documentstyle[psfig,epsf,amssymb,rotating]{aa}
\begin{document}
\thesaurus{02(02.01.2, 02.05.2, 08.06.2, 11.01.2, 11.14.1)}
\title{On the transition to self-gravity\\
in low mass AGN and YSO accretion discs}
 \titlerunning{On the transition to self-gravity in low mass accretion discs}
             \author{Jean-Marc Hur\'e$^{1,2}$}
             \offprints{Jean-Marc.Hure@obspm.fr}
             \institute{$^1$DAEC et UMR 8631 du CNRS, Observatoire de Paris-Meudon, Place Jules Janssen, 92195 Meudon Cedex, France\\
                        $^2$Universit\'e Paris 7 (Denis Diderot), 2 Place Jussieu, 75251 Paris Cedex 05, France}
             \date{Received 31 March 2000 / accepted 11 April 2000}
             \maketitle

\begin{abstract}

The equations governing the vertical structure of a stationary keplerian accretion disc
supporting an Eddington atmosphere are presented. The model is based on the
$\alpha$-prescription for turbulent viscosity (two versions are tested), includes the disc vertical self-gravity, convective transport and turbulent
pressure. We use an accurate equation of state and wide opacity grids
which combine the Rosseland and Planck absorption means through a depth-dependent
weighting function. The numerical method is
based on single side shooting and incorporates algorithms designed
for stiff initial value problems. A few properties of the model are discussed for a circumstellar disc
around a sun-like star and a disc feeding a $10^8$ M$_\odot$ central black hole. Various accretion rates and $\alpha$-parameter values are considered.

We show the strong sensitivity of the disc structure to the viscous energy
deposition towards the vertical axis, specially when entering inside the
self-gravitating part of the disc. The local version of the
$\alpha$-prescription leads to a "singular" behavior which is also predicted by the vertically averaged
model: there is an extremely violent density and surface density runaway, a rapid disc collapse and a temperature plateau. With respect, a much
softer transition is observed with the ``$\alpha
\cal{P}$-formalism''. Turbulent pressure is important only for $\alpha \gtrsim
0.1$. It lowers vertical density gradients, significantly thickens the disc (increases its flaring), tends
to wash out density inversions occurring in the upper layers and pushes the self-gravitating region to
slightly larger radii. Curves localizing the inner edge of the self-gravitating disc as functions
 of the viscosity parameter and accretion rate are given. The lower $\alpha$, the closer to the center the self-gravitating
regime, and the sensitivity to the accretion rate is generally weak, except for
$\alpha \gtrsim 0.1$.

This study suggests that models aiming to describe T-Tauri discs beyond about a few to a few tens astronomical units (depending on the viscosity
parameter) from the central protostar using the $\alpha$-theory should
consider vertical self-gravity, but additional heating mechanisms are
necessary to account for large discs. The Primitive Solar Nebula was probably a bit (if not strongly) self-gravitating at the actual orbit of
 giant planets. In agreement with vertically averaged computations,
 $\alpha$-discs hosted by active galaxies are self-gravitating beyond about a
 thousand Schwarzchild radii. The inferred surface density remains too high to
 lower the accretion time scale as requested to fuel steadily active nuclei for a few hundred millions years. More efficient mechanisms driving accretion are required.
   

\end{abstract}

\section{Introduction}

Viscous discs can acquire large dimensions under the effect of angular
momentum redistribution. This is corroborated by the observation of Young
Stellar Objects (YSO) which reveals the emission of wide gaseous discs with an
outer radius reaching $100 - 1000$ AU (Beckwith et al., 1990; Pudritz et al.,
1996; Duvert et al., 1998; Guilloteau \& Dutrey, 1998; Shepherd \& Kurtz,
1999). Also, discs hosted by Active
Galactic Nuclei (AGN), although yet unresolved |except in the active galaxy
NGC4258 (Herrnstein et al., 1999)| could probably be of much larger size. The
existence of strong similarities between YSOs and AGNs, like jets and outflows
(Falcke, 1998; Frank, 1998), some excesses in the spectral energy distribution at infrared
wavelengths (Adams \& Shu, 1986; Zdziarski, 1986;  Bertout,
1989; Sanders et al., 1989; Voit 1991; Kenyon, Yi \& Hartmann, 1996), a surrounding (dusty)
torus (G\"usten, Chini \& Neckel, 1984; Th\'e \& Molster, 1994; Drinkwater,
Combes \& Wiklind, 1996; Sandqvist, 1999) indicates that the same physical
mechanisms should take place in discs, despite the difference in the central
mass scale. This constitutes an interesting challenge for the disc theory,
especially to understand the transport of angular momentum.

Many models are based on the $\alpha$-theory of thin
discs (Shakura \& Sunyaev, 1973; Pringle, 1981) and assume a vertically
averaged structure (e.g., Collin \& Dumont 1990; Ruden \& Pollack, 1991;
Cannizzo \& Reiff, 1992; Hur\'e et al. 1994a; Artemova et al. 1996; Burderi,
King \& Szuszkiewicz, 1998; Drouart et
al. 1999; Aikawa et al., 1999). A physically more satisfactory approach is the
investigation of the vertical structure in detail, as done for stars. Such a
problem has been discussed by several groups already, in different contexts, with various goals and degrees
of sophistication: at the scale of AGNs (Cannizzo, 1992; Wehrse, St\"orzer \& Shaviv, 1993;
D\"orrer et al., 1996; Siemiginowska, Czerny \& Kostyunin, 1996; Hubeny \&
Hubeny, 1998 and references therein; Sincell \& Krolik, 1997; R\'o\.za\'nska
et al., 1999), in Cataclysmic Variables (CVs) (Smak, 1984; Mineshige \& Osaki, 1983;
Meyer \& Meyer-Hofmeister, 1982; Pojma\'nski, 1986; see Cannizzo, 1993 for a
review; Milsom, Chen \& Taam; 1994; Dubus et al., 1999) and in YSOs (Malbet \&
Bertout, 1991; Bell et al., 1997; D'Alessio et al., 1998) including the
Primitive Solar Nebula (PSN) (Lin \& Papaloizou, 1980; Papaloizou \& Terquem,
1999). Basic considerations show that the outermost regions of (low
mass) discs are expected to be regulated by self-gravity (Goldreich \&
Lynden-Bell, 1965; Shlosman \& Begelman, 1987). This may be the case of AGN
and YSO discs. A few constraints on the disc thickness, mass and rotation
motion at large radii are set by observations (Guilloteau \& Dutrey, 1998; Mundy, Looney \& Welch, 2000;
Herrnstein et al., 1999). So, it appears essential to construct models for
as accurate and realistic as possible, despite the lack of knowledge regarding
turbulent viscosity which causes accretion. To our knowledge, no self-consistent 2D-model accounting for vertical self-gravity in the outermost
parts of discs has been published yet. This is the aim of this paper. The present model includes simultaneously vertical convection, self-gravitation, turbulent pressure, and realistic equation of state (EOS) and
opacities. Of particular interest here is the transition from the
``classical'' disc to the self-gravitating disc which is predicted to be as
close as a few $10^3 -10^4 R_\star$ ($R_\star$ is the radius of the central object)
by vertically averaged models (Ruden \& Pollack, 1991; D'alessio, Calvet \& Hartmann, 1997; Hur\'e, 1998).
The hypothesis of the model and relevant equations are developed in
Sect. 2. Two versions of the $\alpha$-prescription are tested. The ingredients
(EOS and opacities) as well as the numerical method are presented in
Sect. 3. We discuss in Sect. 4 a few properties of the model, namely the effect
of turbulent pressure and depth dependent viscosity, and the position of the
inner edge of the self-gravitating disc, for prototypal circumstellar and AGN
discs, for various accretion rates. An Appendix contains a note concerning
the treatment of convection and a high precision formula fitting the EOS.

\section{Model for the vertical structure: hypothesis and relevant equations}

\subsection{General considerations. Accounting for self-gravity}

A distinctive feature of steady state keplerian accretion discs is the absence
of coupling between the vertical structure and the radial structure
(e.g. Frank, King \& Raine, 1992). This attractive property is undoubtedly an
oversimplification and probably does not match reality. However, it is particularly advantageous from a modeling point of view because any annulus can be treated individually, whatever the state of neighbors, unlike in thick discs where pressure gradients and energy transport in the radial direction gain in importance with respect to their vertical counterparts (Maraschi, Reina \& Treves, 1976; Abramowicz, Calvani \& Nobili, 1980; Abramowicz et al., 1988; Narayan, Madevan \& Quataert, 1998). Note that the keplerian assumption which fixes the rotation law to the value
\begin{equation}
\Omega=\sqrt{\frac{GM}{R^3}}
\end{equation}
where $M$ is the central mass and $R$ the polar radius, requires that the gas
remains confined at altitudes $z$ such as $z^2 \ll R^2$, and small radial
pressure gradients too.

Self-gravity may influence and even dominate the equilibrium structure and
dynamical evolution of almost any kind of disc, not only in massive or thick
discs or tori where effects are global (Bodo \& Curir, 1992; Hashimoto,
Eriguchi, M\"uller, 1995; Boss, 1996; Laughlin \& R\'o\.zyczka, 1996, Masuda,
Nishida \& Eriguchi, 1998), but also in low
mass keplerian discs (Paczy\'nski, 1978; Shlosman \& Begelman, 1987) as soon as the mass density of the accreted gas exceeds
$\sim \Omega^2/4 \pi G$ locally, $\Omega$ being the rotation frequency. In that
latter case which is of interest here, self-gravity increases vertical pressure
gradients and gathers matter closer to the midplane (Sakimoto \& Coroniti,
1981; Shore \& White, 1982; Cannizzo \& Reiff, 1992; Hur\'e et al., 1994; Hur\'e, 1998). So, only the outermost
regions of keplerian discs where $\Omega$ reaches low values can be affected
by vertical self-gravity, except in very special situations (e.g. Sincell \&
Krolik, 1997).

Accounting correctly for
the disc gravity requires the resolution of the Poisson equation (Hunter,
1963; St\"orzer, 1993; Bertin \& Lodato, 1999)
\begin{equation}
\Delta \Phi^{\rm disc}(R,z) = - 4 \pi G \rho(R,z)
\label{eq:fish}
\end{equation}
where $\rho$ is the gas mass density and $\Phi^{\rm disc}$ is the gravitational potential due to the bare disc (the $\phi$-invariance is assumed). This is a very difficult task, in particular because the disc surface has
a non trivial form which is not known a priori and the deviation from sphericity is extreme. As Eq.(\ref{eq:fish}) connects all annuli together, the method of ``independent rings'' no longer applies, unless an iterative scheme in which the potential
would be step by step improved from the actual density field until
convergence (Stahler, 1983). There is however no guarantee that such a scheme
effectively converge and the computational time might be prohibitive (Eriguchi
\& M\"uller, 1991). Here,
we follow another, more simple approach: we adopt the infinite and $R$-homogeneous
slab approximation (Paczy\`nski,
1978; Sakimoto \& Coroniti, 1981; Shore \& White, 1982; Cannizzo \& Reiff,
1992; Liu, Xie \& Ji, 1994; Hur\'e, 1998) which yields the gravity due to the disc
\begin{equation}
g^{\rm disc}_z(R,z) = \frac{\partial \Phi^{\rm disc}}{\partial z} \approx - 4 \pi G \Sigma(R,z),
\label{eq:gdisc}
\end{equation}
where the surface density $\Sigma$ is defined by
\begin{equation}
\Sigma(R,z) = \int_{0}^{z}{\rho(R,z')dz'},
\label{eq:sigma}
\end{equation}
but other assumptions are possible (Mineshige \& Umemura, 1997). This
enables again to investigate the disc structure annulus by annulus, but
introduces a bias: it tends to overestimate (underestimate) self-gravity in
regions of high (respectively low) densities. We expect that the present
approximation gives acceptable results, at least in regions where the surface
density radial gradients remain low with respect to density. Anyway, the error made with respect to a self-consistent model is not known and would be interesting to estimate.

\subsection{The equations for the disc interior}

When the upward transport of heat is treated in the diffusion approximation,
the vertical structure of a keplerian disc, possibly irradiated, can be determined
from the resolution of a system of four first order coupled ordinary
differential equations (ODEs) in between the midplane and the top of the disc
(Pojma\'nski, 1986; Tuchman, Mineshige \& Wheeler, 1990; Cannizzo, 1992; Meyer
\& Meyer-Hofmeister, 1982; Milsom, Chen \& Taam, 1994; Dubus et al.,
1999). The complexity of the problem rises when a multi-frequency radiative
transfer is performed, as required to compare theoretical spectra with
observations and make key predictions (Ross, Fabian \& Mineshige, 1992;
Wehrse, St\"orzer \& Shaviv, 1993; D\"orrer et al., 1996; Sincell \& Krolik,
1997; D'alessio et al. 1998; El-Khoury \& Wickramasinghe, 1998; Hubeny \&
Hubeny, 1998; De Kool \& Wickramasinghe, 1999). The four equations specify
respectively (Frank, King \& Raine, 1992):
\begin{itemize}
\item the pressure gradient $\nabla_z P$ which describes the hydrostatic equilibrium of each fictitious slab
\begin{equation}
\frac{1}{\rho}\frac{d P}{dz} = -\Omega^2 z - 4 \pi G \Sigma \equiv g_z,
\label{eq:eh1}
\end{equation}
\item the heat flux gradient $\nabla_z F$ due to viscous heating
\begin{equation}
\frac{dF}{dz} = \frac{9}{4} \rho \nu \Omega^2,
\label{eq:dfoverdz}
\end{equation}
where $\nu$ is the $z$-dependent viscosity law (see Sect. 2.4),
\item the temperature gradient $\nabla_z T$ which is determined by the heat
  net flux transported upwards through radiation and convection
\begin{equation}
\frac{d T}{d z} = - T \frac{ \nabla }{\lambda_p},
\label{eq:temp}
\end{equation}
where  $\lambda_p \equiv -\frac{dz}{d \ln P}$ is the pressure height scale and
$\nabla \equiv \frac{d \ln T}{d \ln P}$ is the actual gradient (see the
Appendix, Sect. A),
\item the surface density gradient $\nabla_z \Sigma$
\begin{equation}
\frac{d\Sigma}{dz} = \rho
\label{eq:sd2}
\end{equation}
\end{itemize}

Note that this last equation is not relevant when self-gravity is left
 apart (the total surface density of the disc + atmospheres $\Sigma_{\rm t} = 2 \times \Sigma
 (R,\infty)$ can be easily computed a posteriori from Eq.(\ref{eq:sigma}),
 once the mass density distribution is known). The above ODEs must be supplemented by a closure relation, an equation of state. For a mixture of radiation and perfect gas undergoing atomic ionization and molecular dissociation at LTE, the total pressure is linked to the density and temperature of matter by
\begin{equation}
P = \frac{\rho k T }{\mu m_{\rm H}} + \frac{4 \, \sigma}{3 \, c}T^4
\end{equation}
where $\mu m_{\rm H}$ is the mean mass per particle which depends on the
density and temperature (see Sect. 3.1 and the Appendix, Sect. B). The above
expression for radiation pressure is compatible with an optically thick
disc only.

\subsection{Accounting for turbulent pressure in the framework of the $\alpha$-prescription}

Turbulence is the main mechanism driving accretion in discs. It is an extra
source of pressure. According to the standard theory of $\alpha$-discs
(Shakura \& Sunyaev, 1973; Pringle, 1981), the typical velocity of turbulent
eddies can be taken as $\sim \sqrt{\alpha} c_{\rm s}$ if one assumes the
equipartition between length and velocity turbulent scales ($c_{\rm s}$ is the
adiabatic sound speed). So, the turbulent pressure is
\begin{equation}
p_{\rm t} = \alpha \Gamma_1 P
\label{eq:pt}
\end{equation}
where $\Gamma_1= (\frac{d \ln P}{d \ln \rho})_{\rm ad}$ is the first adiabatic
exponent in the total (gas plus radiation) pressure. The effect of turbulent
pressure on the hydrostatic equilibrium is therefore expected if $\alpha$
is not too small, as simulations shall confirm. If we include $p_{\rm t}$ into Eq.(\ref{eq:eh1}) and rewrite
it in terms of a density gradient equation, we find
\begin{equation}
\frac{d \rho}{dz} = -\frac{\rho^2}{P} \left( \Omega^2 z + 4 \, \pi \, G \Sigma \right) \frac{ 1 - \chi_T \nabla}{\chi_\rho \left( 1 + \alpha \Gamma_1 \right)}
\label{eq:eh2}
\end{equation}
where $\chi_T$ and $\chi_\rho$ are respectively the temperature and density
exponents of the total pressure, and we have assumed $\nabla_z \Gamma_1=0$
(this is justified since $0.9 \lesssim \Gamma_1 \lesssim 1.7$, see for
instance the Appendix, Fig.(\ref{fig_coefs})). This expression clearly shows the existence of a density
inversion (that is, a zone where $\nabla_z \rho >0$) each time the actual gradient
satisfies $\chi_T \nabla>1$ (R\'o\.za\'nska et al., 1999). Such an inversion is
likely to occur in radiative pressure dominated layers where $\chi_T$ is the
largest or/and in convectively unstable zones where $\nabla$ may be
large. Note that the inversion still remains if turbulent pressure plays a
role, but with a weaker amplitude.

It is likely that turbulent pressure plays a role, not only on the hydrostatic equilibrium as
considered here, but also on advection of matter and energy both radially and
vertically. We ignore these effects.

\subsection{On depth-dependent viscosity laws}

In vertically averaged disc models, the anomalous viscosity takes locally a
mean value defined from midplane quantities, namely $\nu \equiv \nu^{\rm
  midplane} (= \alpha c^2_{\rm
  s}(0) /\Omega$). Here, we need to specify how $\nu$ varies with the
altitude. This point is critical since our knowledge of turbulent viscosity
remains extremely limited, if not null. There are however two standard ways to
do this within the framework of the $\alpha$-prescription. The most common one is to assume that the shear stress is
proportional to some pressure (the so-called "$\alpha {\cal P}$-formalism"; e.g. Cannizzo, 1992). It leads to a viscosity law
\begin{equation}
\nu_1 = \frac{2 \, \alpha \cal P}{3 \, \Omega \rho } \equiv \nu_1(R,z)
\label{eq:nu1}
\end{equation}
where ${\cal P}$ can be either gas pressure, or total pressure or some combination of the two (Siemiginowska, Czerny \& Kostyunin, 1996; Artemova et al., 1996; Hameury et al., 1998; Papaloizou \& Terquem, 1999), with different consequences on the disc stability (Camenzind, Demole \& Straumann 1986; Clarke, 1988). The other way uses the local version of the $\alpha$-prescription (Meyer \& Meyer-Hofmeister, 1982)
\begin{equation}
\nu_2 = \alpha c_{\rm s} \bar{\lambda}_p \equiv \nu_2(R,z)
\label{eq:nu2}
\end{equation}
where $\bar{\lambda}_p$ is the reduced pressure height scale (see the
Appendix, Sect. A). Whatever the expression the authors adopt, the
$\alpha$-parameter is most often assigned to a fixed value in a disc. However, theoretical arguments, numerical
simulations as well as observational constraints indicate that
$\alpha$ should vary, not only with the radius (Shakura \& Sunyaev, 1973;
Cannizzo, 1993; Lasota \& Hameury, 1998), but also with the altitude via the temperature or density, or
other quantities (Brandenburg, 1998). Strictly speaking, the need for strong
variations of the $\alpha$-parameter means the failure of the $\alpha$-viscosity model.

Since $\nu_1$ and $\nu_2$ are formally different, even with the same
$\alpha$-parameter, they lead to different vertical structures and
consequently to different discs, as we shall see below. Besides, the
correspondence between the two laws (a multiplying
factor $\frac{3}{2} \Omega \lambda_p \Gamma_1/c_{\rm s}$ from $\nu_1$ to
$\nu_2$, if ${\cal P}\equiv P$) is
much more subtile than shifting $\alpha$: in general, $\Omega \lambda_p \ne
c_{\rm s}$ at any altitude (this is specially true at the equatorial plane where
$\lambda_{\rm p}\rightarrow \infty$; see Sect. 4.3). The volumic energy production associated to $\nu_2$ is
\begin{equation}
\frac{dF}{dz} = \frac{9}{4} \Omega^2  \lambda_p \alpha \sqrt{\Gamma_1 P \rho},
\label{eq:dfoverdz2}
\end{equation}
 Note that both Eqs.(\ref{eq:nu1}) and
(\ref{eq:nu2}) satisfy $\frac{d \nu}{dz} < 0$ above the equatorial plane,
meaning that the gas is accreted much faster at the equatorial plane than at
the top of the disc. So, these laws implicitly suggest the existence a plan parallel shear which might be able to trigger turbulence and to expand turbulent eddies in the radial direction. 

There are still no physical arguments to decide if $\nu_2$ is
better than $\nu_1$. That is why we use both expressions in the following (in particular, for $\nu_1$ we take ${\cal P} \equiv P$). In
fact, a wide class of functions $\nu(R,z)$ should be considered and their effects compared. It is possible that turbulent viscosity shows a weaker dependence with $z$
than considered so far and even does not depend on the altitude at all. This
is precisely the case with the $\beta$-viscosity prescription which is suggested by
laboratory experiments (Pringle \& Rees, 1972; Lynden-Bell \& Pringle, 1974; Richard \& Zahn, 1999; Duschl, Biermann \& Strittmatter, 2000; Hur\'e, Richard \& Zahn, 2000).

\subsection{Equations for the Eddington atmosphere}

In the Eddington approximation, the structure of the atmosphere is governed by three (only two in the absence of self-gravity) first orders ODEs. Two of these (namely Eqs.(\ref{eq:eh1}) and (\ref{eq:sd2})) are basically the same as for the disc interior, and the third one controls the optical depth $\tau$ in the atmosphere
\begin{equation}
\frac{d\tau}{dz} = - \kappa \rho
\label{eq:tau}
\end{equation}
where $\kappa$ is a grey absorption coefficient. In order to prevents any
temperature runaway which would lead to the formation of a hot corona (Shaviv \& Wehrse, 1991; R\'o\.za\'nska et al., 1999), we assume that there is no
viscous energy generation in this layer and no active turbulence. It means that the atmosphere is not accreted. We are aware that the
validity of all these approximations, including the Eddington approximation,
is easily open to criticism. The temperature within the atmosphere follows the law (Mihalas, 1978)
\begin{equation}
T^4=\frac{3}{4} T^4_{\rm eff} \left(\tau +\frac{2}{3}\right)
\label{eq:tdistrib}
\end{equation}
 where $\tau \le \frac{2}{3}$ and $T_{\rm eff}$ is the effective temperature
 which is fixed by the outgoing flux due to internal and external heating
 sources, at the photosphere's base. Actually, the global effect of the disc irradiation can
 be taken into account at this level by a suitable definition of $T_{\rm eff}$
 (Kenyon \& Hartmann, 1987; Hubeny, 1990; Robinson, Marsh \& Smak, 1993; Dubus et al., 1999). Wee see from Eqs.(\ref{eq:eh1}), (\ref{eq:tau}) and (\ref{eq:tdistrib}) that the density gradient is 
\begin{equation}
\frac{d \rho}{dz} = -\frac{\rho^2}{\chi_\rho P} \left( \Omega^2 z + 4 \, \pi \, G \Sigma \right) + \frac{\chi_T \kappa \rho}{4\, \chi_\rho \left(\tau+\frac{2}{3} \right)}
\label{eq:eh3}
\end{equation}
and may also become positive. As suggested above, we do not take into account turbulent
pressure in this layer. This induces a discontinuity in the density (or gas
pressure) gradient at the altitude where the disc and its atmosphere join together, but not in the density itself.

\subsection{Boundary conditions and matching conditions at the disc/atmosphere
  interface}

There are two problems to solve simultaneously: the structure of the disc interior
from Eqs.(\ref{eq:dfoverdz})(\ref{eq:temp})(\ref{eq:sd2}) and (\ref{eq:eh2})
and the structure of its atmosphere from Eqs.(\ref{eq:sd2})(\ref{eq:tau}) and
(\ref{eq:eh3}). These differential
equations are subject to Dirichlet boundary conditions. At the midplane, the
symmetry imposes $\Sigma=0$ and $F=0$ like for a star. At the top atmosphere located at
$z=H$ (to be determined), conditions are $P = P_{\rm amb}$ the ambient
pressure (which may include the pressure due to external illumination), $\tau
= 0$ and $\Sigma=\frac{1}{2}\Sigma_{\rm t}$, half the total surface density of
the disc and atmosphere. Some authors specify the density (for instance $\rho
\rightarrow 0$) instead of the
pressure at the boundary (e.g. Pojma\'nski, 1986; D\"orrer et al. 1993; El-Khoury \& Wickramasinghe, 1999).  We choose a value typical of
the interstellar medium, namely $P_{\rm amb}/k = 10^5$ K.cm$^{-3}$ (Duley \& Williams, 1988),
significantly smaller than in D'alessio et al. (1998). At the base of the atmosphere located at $z=h$ (to be
determined too), the matching conditions are $\tau=\frac{2}{3}$ and $F=\sigma
T^4_{\rm eff}$ (see Sect. 4.1).

It is worthy of note that boundary conditions at the disc surface have a very
weak effect on midplane quantities as long as the disc is optically thick and
has a large surface density, locally. Such an insensitivity is well known in stellar structure computations (Kippenhahn \& Weigert, 1990). Let us quote also that, contrary to a general
 belief, models that do not consider the atmosphere and use $P = - \frac{2}{3}
 \frac{g_z}{\kappa}$ as a boundary condition at $z=h$ always violate
 the Eddington approximation since this relation does not guarantee that the atmosphere has the right optical thickness (i.e. $\frac{2}{3}$). Further, the approximation
\begin{equation}
\int_{P(\infty)}^{P(h)}{\frac{\kappa}{g_z} dP} \approx \left( \frac{\kappa}{g_z} P \right)_{z=h}
\label{eq:bcpexact}
\end{equation}
becomes bad in regions where the surface density of the atmosphere is comparable
to that of the disc, when the mean absorption within the atmosphere is low, or
when the disc is not optically very thick. For these reasons, we sustain the
idea that the use of the Eddington approximation has sense only if the
structure of the atmosphere is computed together with that of the disc
interior.

\section{Ingredients and computational method}

\subsection{Equation of state and opacities for a cosmic gas}

  The physical problem depends on a certain amount of thermodynamical and
  radiative data. For the present application, the gas has cosmic abundances
  and is subject to atomic ionizations and molecular dissociations. Local Thermodynamic Equilibrium (LTE) is assumed (Hur\'e et al., 1994b). The mean
  mass per particle $\mu m _{\rm H}$, coefficients $\chi_T$ and $\chi_\rho$,
  adiabatic gradient $\nabla_{\rm ad}$, heat capacity $c_p$ (needed to treat
   convective transport) and adiabatic exponent $\Gamma_1$ have been computed
  accurately from chemical abundances at thermal equilibrium (Hur\'e,
  1998). Grids of data with $\rho$ and $T$ as the inputs
  have been generated. Interpolations
  between mesh points are performed with bicubic splines.
We have computed a high precision
  analytical expression fitting $\mu$ (and subsequently coefficients $\chi_T$
  and $\chi_\rho$) as functions of the temperature and density with an
  accuracy less than 4$\%$ relative to raw data in the case of a zero
  metallicity gas (see the Appendix, Sect. B1). This fit can be used for a
  cosmic gas as well without producing important errors (since metals in a
  cosmic mix have very low abundances relative to hydrogen, with almost no  influence on the EOS and related quantities).

Opacity is probably the most important ingredient of the model and the choice for the grey absorption coefficient $\kappa$ is critical, specially near the surface (Mihalas, 1978). Here, we take (Hameury et al., 1998)
\begin{equation}
\kappa =  \theta \kappa_{\rm R} + (1-\theta)\kappa_{\rm P}
\label{eq:kgrey}
\end{equation}
where $\kappa_{\rm R}$ and $\kappa_{\rm P}$ are the Rosseland and Planck means respectively, and $\theta$ is a function which varies continuously from 0 to 1 as the optical depth increases from 0 to infinity. A suitable choice is
\begin{equation}
\theta_m(\tau) = \frac{1}{1+\tau^m}
\end{equation}
where the index $m$ sets the stiffness of the transition from
$\kappa_{\rm P}$ to $\kappa_{\rm R}$ ($m=1$ is taken in the following
applications). The $\theta$-function is arbitrary and should strongly
influence physical quantities at the disc surface. Also, we are aware that the best grey coefficient is probably neither given by $\kappa_{\rm R}$, nor by $\kappa_{\rm P}$, nor by Eq.(\ref{eq:kgrey}). A flux weighted opacity mean could be better (Mihalas, 1978; Hubeny, 1990).

\begin{figure}
\psfig{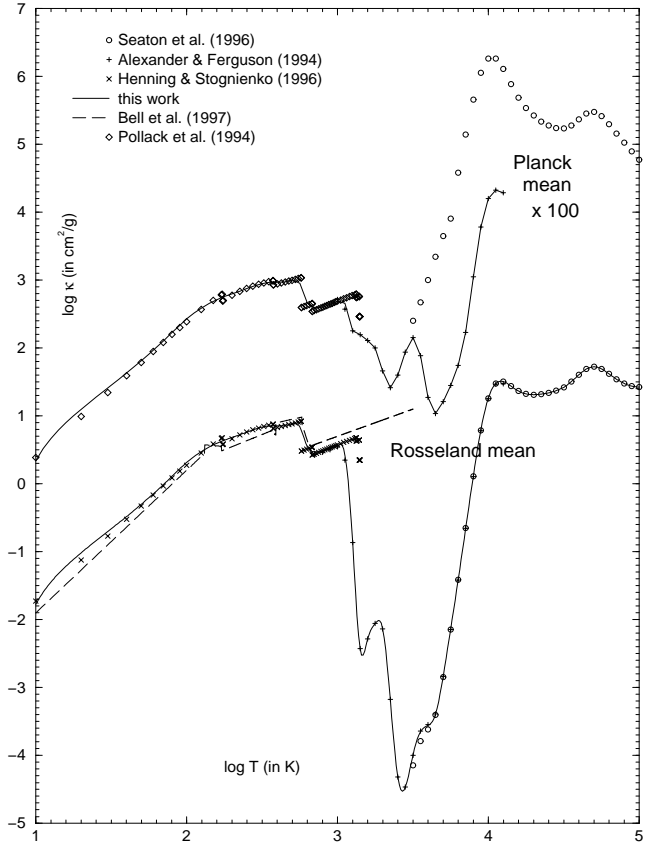}
\caption{Rosseland and Planck absorption coefficients versus the temperature
  from various sources for hydrogen and metal mass fractions $X=0.70$ and
  $Z=0.02$ respectively and $\rho=10^{-15} \times
  T^3$ g.cm$^{-3}$ (i.e. $\log R=-3$ in the OP format; see Seaton
  et al., 1994). Planck means have been shifted upwards for clarity.}
\label{fig_opac}
\end{figure}

Tables of Rosseland and Planck means have been taken from various sources
(Pollack et al., 1994; Hur\'e et al., 1994b; Alexander \& Fergusson, 1994; Henning \&
Stognienko, 1996, Seaton et al., 1994). But Planck absorption means published
by the Opacity Project (OP) (Seaton et al., 1996) have not be selected due to
apparently large errors of unknown origin (Alexander, 1998; Zeippen,
1999). Two grids of data have been generated after passing through filters
(mainly running averages) to smooth out discontinuities. Finally, opacity
grids cover the temperature range $10-10^8$ K and extend from extremely high
to extremely low densities where the ideal gas and LTE assumptions are expected to fail. An example of Rosseland and Planck means versus the temperature is displayed in Fig.(\ref{fig_opac}). As done for the equation of state and related quantities, opacity values between mesh points are interpolated using bicubic splines. Working with analytical opacities may present some advantages (Burgers \& Lamers, 1989; Collin \& Dumont, 1990; Bell \& Lin, 1994), but they can also produce numerical instabilities in the integration of the disc structure equations since they mostly consist in continuous but non derivable piecewise functions.

\subsection{Computational method}

The altitude of the top atmosphere being not known a priori, many numerical methods for
solving Two Boundary Value Problems (TBVPs) are almost unusable in the actual
situation. It is however possible to apply variable changes, namely $\nabla_z
\rightarrow \nabla_\tau$ in the atmosphere and $\nabla_z \rightarrow \nabla_F$
in the disc interior (Cannizzo \& Cameron, 1988) in order to recover a common TBVP with fixed boundaries (e.g. $\frac{F}{\sigma T_{\rm eff}^4} \in [0, 1]$), but this drastically increases vertical gradients near the surface and make the problem more
unstable from a numerical point of view. Some groups work with relaxation
algorithms (Cannizzo, 1992; Milsom, Chen \& Taam, 1994; D'alessio et al.,
1998; Hameury et al., 1998) which require (good) vertical profiles as starting
guesses and are known to be rapidly converging methods. However, zones with steep gradients generally need a local mesh refinement. Other authors prefer straight integrators and algorithms based on shooting methods (Lin \& Papaloizou, 1980; Meyer \& Meyer-Hofmeister, 1982; Mineshige \& Osaki, 1983; Smak, 1984; Mineshige, Tuchman \& Wheeler, 1990; R\'o\.za\'nska et al., 1998; Papaloizou \& Terquem, 1999) typical of Initial Values Problems (IVPs), with only a few quantities to guess but with possible troubles regarding the precision.

In the present case, we have written a Fortran code named {\tt VS$^{\tt
    3}$KAD} which performs the integration of the disc and atmosphere
equations using single side shooting and numerical routines specially designed for
    stiff IVPs. Integration steps are variable, internally to the routines and the
accuracy can be as small as the machine precision. In its present version, the
    code works with the following variables: $\ln
T$, $\ln \rho$, $\frac{F}{\sigma T_{\rm eff}^4}$, $\frac{2 \Sigma}{\Sigma_{\rm
    t}}$ and $\tau$, but more clever choices are possible,
   specially to reduce artificially the stiffness of the equations. We have noticed that the
    computational time is considerably smaller than with Runge-Kutta
    algorithms, even with a variable step. Like in many stiff problems, the
    numerical integration may turn out to be non conservative: integrating
    from the top atmosphere down to the midplane can give a solution which, in
    the surface neighborhood, can be very different than that obtained when integrating in
    the opposite direction (Mineshige, Tuchman \& Wheeler,
    1990; Press et al., 1992; De Kool \& Wickramasinghe, 1999). This has been
    observed in the present case. It reminds the influence of both boundary conditions and underlying methodology. The best stability and reliability of the results are obtained by starting from the boundary layer where almost all the stiffness of the problem is concentrated.

In practical, given a central mass $M$, accretion rate $\dot{M}$,
$\alpha$-parameter, radius $R$ and total energy deposition $\sigma T^4_{\rm
  eff}$ at the bottom atmosphere (see Sect. 4.1), the computation starts at an arbitrary
altitude $z=H^{(0)}$ (the top atmosphere) where the surface density
$\Sigma^{(0)}(H^{(0)})$ is guessed. The integration then proceeds
downwards. Once the bottom atmosphere is reached, at an altitude $z=h^{(0)}$, the equations for the disc interior are integrated down to the midplane where the net flux and surface density generally differ from zero. By successive iterations on $H^{(i)}$ and $\Sigma(H^{(i)})$ |performed with a Newton-Raphson method|, quantities $F^{(i)}(0)$ and $\Sigma^{(i)}(0)$ can be driven to very small values. The problem has converged after $n$ iterations, when simultaneously $F^{(n)}(0) \approx 0$ and $\Sigma^{(n)}(0) \approx 0$ with the requested precisions
\begin{equation}
-\epsilon^{F}_- \le \frac{F^{(n)}(0)}{\sigma T_{\rm eff}^4} \le \epsilon^{F}_+,
\end{equation}
\begin{equation}
-\epsilon^{\Sigma}_- \le \frac{2 \, \Sigma^{(n)}(0)}{\Sigma_{\rm t}} \le \epsilon^{\Sigma}_+
\end{equation}
and
\begin{equation}
-\epsilon^{H}_- \le 1- \frac{ H^{(n)} } { H^{(n-1)} } \le \epsilon^{H}_+,
\end{equation}
where $\epsilon^{F}_+$, $\epsilon^{F}_+$, $\epsilon^{\Sigma}_+$,
$\epsilon^{\Sigma}_-$, $\epsilon^{H}_-$ and $\epsilon^{H}_+$ are very small, positive values. Since our treatment of self-gravity is very approximate, it is legitimate to allow a much lower (but
sufficient) precision on the surface density (this accelerates convergence and
lowers the computational time). On average, it takes a few milliseconds CPU on
a single user personal computer and $n \lesssim 25$ with $\epsilon^{F}_0=
\epsilon^{H}_0=(\epsilon_0^{\Sigma})^2= 10^{-10}$ where $\epsilon^{F}_0 = \rm Sup
\left( \epsilon^{F}_-, \epsilon^{F}_+\right)$, $\epsilon^{\Sigma}_0 =
\rm Sup \left( \epsilon^{\Sigma}_-, \epsilon^{\Sigma}_+\right)$ and $\epsilon^{H}_0
= \rm Sup \left( \epsilon^{H}_-, \epsilon^{H}_+\right)$. This precision is
sufficient in most cases. When self-gravity is neglected,
Eq.(\ref{eq:sd2}) becomes obsolete. The iteration is on the flux only and the
code then runs much faster (by a factor $\sim 5$), with $n \lesssim 6$.

The code has been tested and compared in many situations
(e.g. without self-gravity, without turbulent pressure, with and without
external irradiation) with reference models (R\'o\.za\'nska, 1998; Hameury et
al., 1998) and appears to behave very well\footnote{The author plans to make
  the executable available to the community. In the meanwhile, vertical
  structure computations may be performed on special request.}. It
automatically scales to the central mass, and is able to model any kind of
disc: AGN discs, CV discs, YSO discs and sub-nebulae as well. A simulation of
D/H enrichments in the PSN with this code is currently under way (Hersant, Gautier, Hur\'e, 2000).

\section{A few applications of the model}

\subsection{The hypothesis of a non irradiated disc}

Vertically averaged disc models show that the non self-gravitating and gas pressure
dominated parts of $\alpha$-discs have an aspect ratio $\frac{h}{R} \sim 0.001 - 0.1$, depending on the central mass and accretion rate
mainly. There is a slight flaring (i.e. $ \frac{d \ln h }{ d \ln R} \gtrsim 1$)
which eventually may vanish in the outermost regions (e.g. Ruden \& Pollack,
1991; Bell et al., 1997). The existence of a flaring inner disc is supported by observations: the
spectrum of T-Tauri stars and AGN contains a prominent infrared component which is currently interpreted as
light re-processing in the superficial layers of the disc (Adams \& Shu, 1986;
Voit, 1991; Chiang \& Goldreich, 1997). It is possible that
self-irradiation also plays a role (Fukue, 1992). For circumstellar discs, the flaring angle of bare $\alpha$-disc is too low to
enhance the infrared spectral component to the observed level (Kenyon \& Hartmann, 1987). And the heating by the central star should not produce a
significant increase of the flaring (e.g. D'alessio et al., 1999). But the long
wavelength emission can efficiently be boosted if the disc gets into a warped configuration (Terquem \& Bertout, 1993; Miyoshi et al., 1995; Bachev, 1999).

Beyond a certain radius, self-gravity becomes
important and reduces the disc thickness such that $\frac{d \ln h }{ d \ln R} < 0$ (Sakimoto \& Coroniti, 1981; Shore \&
White, 1982; Cannizzo \& Reiff, 1992; Liu, Xie \& Ji, 1994; Hur\'e et al., 1994a; Hur\'e, 1998). As we shall see
below, the same effect is observed from vertical structure computations. It means that outer parts should not receive directly light emitted
at the center, contrary to the inner parts. Note that a disc is generally not isolated but embedded into a "warm"
environment which may in turn heat it up. Discs in YSO are surrounded by an envelope of gas and dust resulting from the cloud core collapse (Mundy, Looney \& Welch, 2000; Chick \&
Cassen, 1997; D'Alessio, Calvet \& Hartmann, 1997). In the case of AGN, clouds
moving above the disc (BLR clouds) can scatter light back onto it (Shields 1977; Collin-Souffrin, 1987;
Osterbrock, 1993). It follows that, even outer regions which are not directly exposed to
the luminous central source can be substantially irradiated. The disc response to irradiation is a complex problem to
solve self-consistently since it depends on many parameters (size and location
of the ionizing source(s), shape of the ionizing spectrum, intensity of
irradiation, disc flaring angle, surface albedo, disc optical thickness, gas
metalicity, etc.) which are neither well known, nor well constrained by
current observations. For discs having a large total surface density, mostly
the superficial layers are heated up and ionized, and the
midplane is essentially not affected (Sincell \& Krolik, 1997; Collin \& Hur\'e, 1999;
Nayakshin, Kazanas \& Kallman, 1999; Igea \& Glassgold, 1999). Deep structural
changes (for instance, a temperature inversion or an iso-thermalization) may
however occur in some cases (D'alessio et al. 1998). Although discs should experience some external heating, even in their
outermost (self-gravitating) parts, we do not consider irradiation in this paper, for simplicity. So, considering internal viscous heating
only, the disc effective temperature, far from the central object, is given by
\begin{equation}
T^4_{\rm eff} = \frac{3 }{8 \, \pi \sigma} \Omega^2 \dot{M}
\end{equation}
where  $\dot{M}$ is the mass accretion rate.

In the following, we apply the code to the computation of the vertical
structure of two different systems subject to vertical self-gravity:
circumstellar discs and AGN discs. We restrict to central masses of $1$
M$_\odot$ and $10^8$ M$_\odot$ respectively, and we show on a few examples how self-gravity and turbulent pressure
which are the main specificities of this model affect the disc
structure. Computations are performed discarding any kind of instabilities
 the disc could undergo. These are stopped when the temperature at the top
atmosphere (i.e. $T(H)=\left(\frac{1}{2}\right)^{1/4} \, T_{\rm eff}$) attains 10 K, for both physical and practical reasons.

\begin{figure}
\psfig{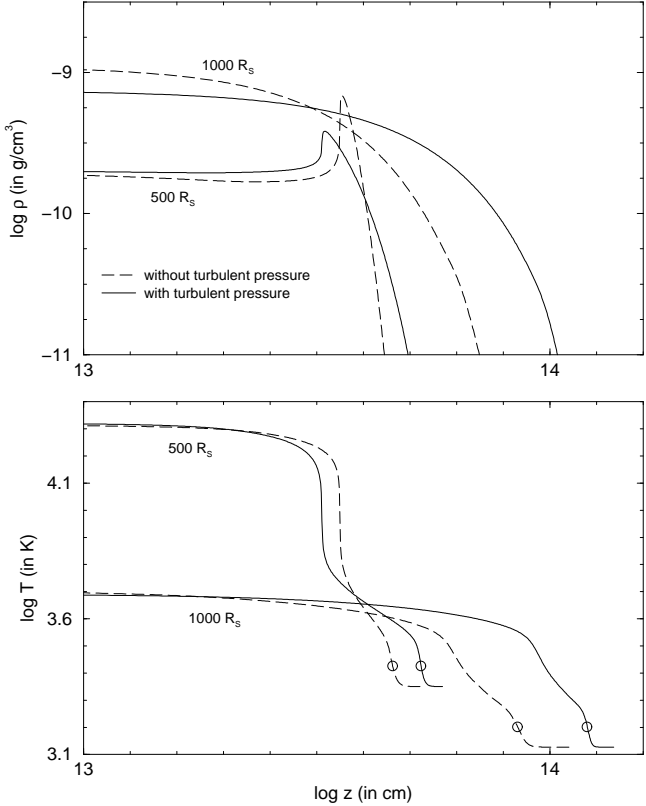}
\caption{Effect of turbulent pressure on the distribution of the density ({\it
    top}) and temperature ({\it bottom}) with the altitude $z$ in an AGN disc
    with $M=10^8$ M$_\odot$, $\dot{M}=10^{-1}$ M$_\odot$/yr and $\alpha=1$ at
    $500$ and $1000 \, R_{\rm S}$ from the center ($R_{\rm S}$ is the
    Schwarzchild radius of the black hole). The position of the bottom
    atmosphere is marked by a circle.}
\label{fig:effecttp}
\end{figure}

\subsection{Effect of turbulent pressure on the vertical temperature and
    density stratification: an example}

To see clearly the influence of turbulent pressure (see Eq.(\ref{eq:pt})), we have chosen a simulation with
$\alpha=1$ which  corresponds to the very upper limit regarding supersonic
turbulence. In addition, convection and self-gravitation are turned off. Under
these circumstances, we have computed the vertical structure with viscosity
law $\nu_1$ for an disc around a massive black hole with $M=10^8$ M$_\odot$, at two different radii: $R=500 \, R_{\rm S}$ ($R_{\rm S}=2GM/c^2$ is the Schwarzchild radius of the black hole) and $R=1000
\, R_{\rm S}$. The accretion rate is $\dot{M}=10^{-1}$ M$_\odot$/yr. The temperature and density profiles so obtained are plotted in Fig.(\ref{fig:effecttp}). We
see that turbulent pressure makes the disc thicker, as expected
intuitively. Actually, according to Eq.(\ref{eq:eh2}), the density gradient is
lowered with turbulent pressure and the transition from the ambient
medium to the disc interior is softer. The flux gradient depending on the density to a positive power |in the absence of convection at
least|, a larger integration range is needed to obtain a
zero net flux at the midplane, hence a thicker disc. The effect is specially visible at $R=1000
\, R_{\rm S}$, with a $41 \%$ disc thickening (measured on $h$; we have $33 \%$ on $H$). The disc
flaring therefore increases by the same relative quantity. Since we have not
explored the whole parameter space, it is probably easy
to find cases where the effect is more important. The total surface density is
almost unchanged (a $2 \%$ increase only). At $500 \, R_{\rm S}$, the disc
thickening is less important, about $15 \%$. Note in this example the presence of a
density inversion which occurs well below the base of
the atmosphere. It tends to be washed out by turbulent pressure, as argued in Sect.2.3. 

The way turbulent pressure acts on both the temperature and density at the
midplane temperature is however less straightforward and depends intimately on
the vertical stratification. For a given effective temperature, a disc with a
larger surface density would theoretically have a hotter core. But this
happens only in the calculation at $500 \, R_{\rm S}$, and the effect is
very minor.

We have checked that the conclusions derived here-above are qualitatively unchanged using
$\nu_2$ (see Sect. 4.5 for another effect of turbulent pressure). We will not discuss the case of circumstellar discs because values of the $\alpha$-parameter
currently adopted for these systems are rather very low (typically $\alpha
\sim 10^{-3} - 10^{-2}$; D'Alessio, Calvet \& Hartmann, 1997) and so, no effect of
turbulent pressure is indeed observable. This has been verified.

\subsection{Effect of the viscosity law: $\nu_1$ versus $\nu_2$}

\begin{figure}
\psfig{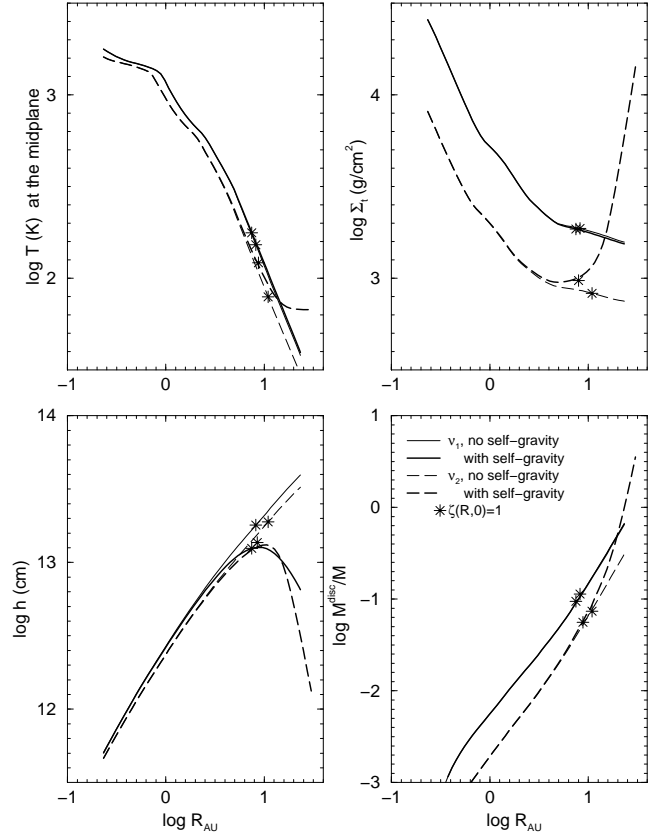}
\caption{Midplane temperature, total surface density, geometrical thickness and disc
  mass versus the radius (in AU) computed  with $\nu_1$
  ({\it solid lines}) and $\nu_2$ ({\it dashed lines}) for a disc with
  $\dot{M}=10^{-7}$ M$_\odot$/yr and $\alpha=10^{-3}$ around a one solar mass
  star.}
\label{fig:rprofyso}
\end{figure}

In this paragraph, we show a few differences between $\nu_1$ and
$\nu_2$. Convection is taken into account and turbulent pressure is left
aside. The midplane temperature,
total surface density, disc thickness and disc mass are plotted versus 
radius in Fig.(\ref{fig:rprofyso}) for $M=1$ M$_\odot$, $\dot{M}=10^{-7}$ M$_\odot$/yr and
$\alpha=10^{-3}$. These
parameter values are typical of discs around T-Tauri stars (D'Alessio, Calvet \& Hartmann, 1997). The disc
mass $M^{\rm disc}$ is
\begin{equation}
M^{\rm disc}(R) = 2 \pi \int^R_{R_{\rm in}}{ R' \Sigma_{\rm t}(R') dR'} \ne \pi \Sigma_{\rm t} R^2
\end{equation}
where $R_{\rm in}$ is the inner edge of the disc (unimportant as long as $R
\gg R_{\rm in}$). Results computed without self-gravity are also shown in
comparison. We see that, in the non self-gravitating region, differences on
the geometrical thickness and midplane temperature are small, the disc being
slightly thicker and hotter with $\nu_1$ than with $\nu_2$. The most important
effect is on the surface density and consequently on the disc mass: the
$\nu_1$-prescription leads to slightly more massive disc than the
$\nu_2$-prescription (by a factor $\sim 2$ here).

We see that the disc is definitely affected by self-gravity from
approximately $7$ AU whatever the viscosity law (in fact, a little bit less with $\nu_1$ which
agrees with the fact that that prescription leads to more massive discs, as long as
$M^{\rm disc} \lesssim M$). Let us remind that the importance of self-gravity
can be measured by the quantity (see Eq.(\ref{eq:eh1}))
\begin{equation}
\zeta(R,z) = \frac{4 \pi G \Sigma(R,z)}{\Omega^2 z}
\label{eq:zeta}
\end{equation}
which remains finite at $z=0$ as a Taylor expansion shows. It follows that a "natural" definition for the limit between the standard disc and the
self-gravitating disc can be $\zeta=1$ at the midplane (if $\zeta>1$
then the disc contribution to gravity exceeds that due to the central object,
vertically).

On the example, the disc thickness goes through
a maximum near 10 AU for both viscosity laws. Beyond this radius, the disc becomes thinner
and thinner. Regions located farther away can therefore not
intercept photons emitted at the center. Interestingly, the temperature and the surface density show very
different behaviors. With law $\nu_1$, the temperature decreases monotonically
as $R$ increases, almost as in the absence of self-gravity. So does the total
surface density. The midplane density, not shown on graphs, gently increases
with the radius. Note that $\Sigma_{\rm t}$ (and $M^{\rm disc}$) are
surprisingly not
affected by self-gravity. The origin of this insensitivity has not been identified
at the time being. Conversely, with law $\nu_2$, the surface density and the
density violently increase with the radius. The disc thickness falls in also
very rapidly. The temperature reaches a plateau. In fact, this somewhat "singular" behavior is predicted by the
vertically averaged model (Shlosman \& Belgelman, 1987; Hur\'e et al. 1994a;
Hur\'e, 1998; see also Duschl, Biermann \& Strittmatter, 2000). In particular,
the temperature  $T_{\rm sg}$ at the plateau is mainly fixed by the ratio $\dot{M}/\alpha$, namely
\begin{eqnarray}
T_{\rm sg} & = &\frac{\mu m_{\mathrm H}}{k} \left(\frac{4 \, G^2 \dot{M}^2}{9 \, \alpha^2}\right)^{1/3}
\nonumber
\\
& \simeq & 24100 \; \mu \left( \frac{\dot{M}}{1 \; {\rm M}_\odot{\rm /yr}} \right)^{2/3} \alpha^{-2/3} \qquad {\rm K}
\label{PSGtg}
\end{eqnarray}
For $\dot{M}/\alpha=10^{-4}$ M$_\odot$/yr as in Fig.(\ref{fig:rprofyso}),
Eq.(\ref{PSGtg}) yields $T_{\rm sg} \simeq 120$ K (assuming $\mu =
2.33$). Although this value is derived from the one zone model, it is in good agreement with the vertical structure computation
which gives a plateau at about $70$ K. Note that the drastic increase
of the surface density produces an important increase of the disc mass which
even exceeds the value obtained with law $\nu_1$. When $M^{\rm disc} \gtrsim
M$ (this occurs at $23 - 28$ AU on the example), the disc becomes very self-gravitating, probably gravitationally unstable (Goldreich \& Lynden-Bell, 1965;
Shu et al., 1990) and should therefore not be well described with current
viscosities (e.g. Lin \& Pringle, 1987) nor by steady state solutions (Goldreich \&
Lynden-Bell, 1965). One
reaches here the limit of the model. It is however interesting to note that,
asymptotically, there is either a hot very dense solution or a cold more diffuse
solution (see Paczy\'nski, 1978).

\begin{figure}
\psfig{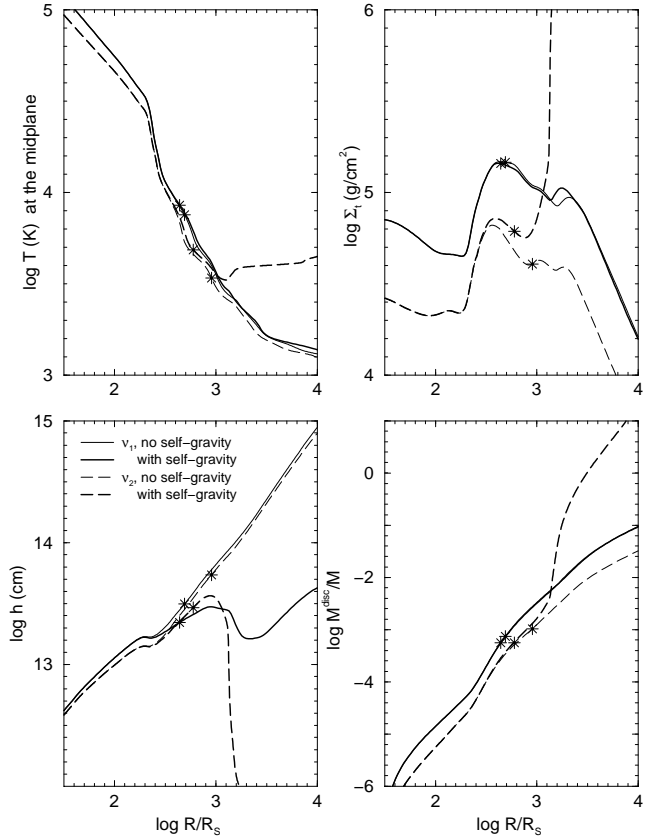}
\caption{Same legend as for Fig.(\ref{fig:rprofyso}) but for an AGN disc with
  $M=10^8$ M$_\odot$, $\dot{M}=10^{-2}$ M$_\odot$/yr and $\alpha=0.1$ ($R_{\rm
  S}$ is the Schwarzschild radius of the black hole).}
\label{fig:rprofagn}
\end{figure}

The same quantities obtained under the same conditions but for an AGN disc with $M=10^8$ M$_\odot$,
$\dot{M}=10^{-2}$ M$_\odot$/yr and $\alpha=0.1$ are
displayed in Fig.(\ref{fig:rprofagn}). Globally, we find similar
trends. The singular behavior observed in the previous example with $\nu_2$
 inside the self-gravitating regime seems even stronger. Self-gravitation
 becomes important from about $500 \, R_{\rm S}$. The midplane temperature
stabilizes radially at $\simeq 3900$ K. This value again compares quite well
with Eq.(\ref{PSGtg}) which gives $6600$ K (assuming $\mu=1.27$). The presence of extremely steep surface density and density radial gradients probably means that the assumption made
on the disc potential (see Sect. 2.1) is no longer valid. This would be
worthwhile to check. The physical picture there is then that of a disc surrounded
by a very dense ring ($\Sigma$ increases by more than one order of magnitude
over a relative length $\frac{\Delta R}{R} \sim 15 \%$) which contains almost all the disc mass and
angular momentum. When $M^{\rm disc} \gtrsim M$ (beyond a few $10^3 \, R_{\rm
S}$ for law $\nu_2$), effects of self-gravity are expected to be global, with a change in the rotation law.

All these results require a few comments. First, it is important to note that
the effect of self-gravity is slightly underestimated when it is neglected in the computations. This is specially
true with law $\nu_2$. Besides, self-gravity becomes important before $\zeta$ reaches unity, like in the one
zone model (Hur\'e, 1998). At the inner edge of the self-gravitating disc
(that is at $\zeta=1$ following our definition), the disc mass is much less than the central
mass: we find $\frac{M^{\rm disc}}{M} \sim 10^{-2}-10^{-1}$ (depending on the
viscosity law) in the YSO case, and $\frac{M^{\rm disc}}{M} \sim 10^{-3}$ in the AGN
case. This suggests that low mass discs must not be automatically classified
as non self-gravitating discs as often asserted. In particular, it is not excluded that T-Tauri
discs we observe be subject to vertical self-gravity, despite their
relatively low mass (Beckwith et al., 1990).

\begin{figure}
\psfig{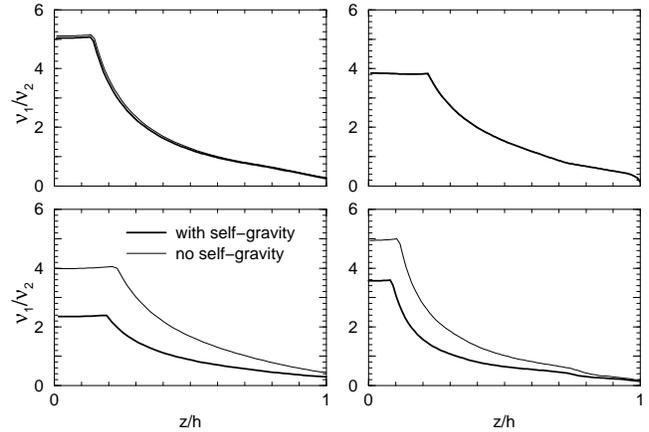}
\caption[]{Ratio $\nu_1/\nu_2$ versus $z/h$ using $\nu_1$ for an YSO disc (same parameter values as for Fig.(\ref{fig:rprofyso})) at 1 AU ({\it top left}) and 10 AU ({\it bottom left}), and for an AGN disc (same parameter values as for Fig.(\ref{fig:rprofagn})) at $100 \, R_{\rm S}$ ({\it top right}) and $500 \, R_{\rm S}$ ({\it bottom right}).}
\label{fig:deltanu}
\end{figure}

It has been stated in Sect. 2.4 that the relation between $\nu_1$ and $\nu_2$
is not trivial. In the non self-gravitating parts however, midplane
temperatures $T(0)$ show a striking quasi-parallel variation with the radius
in logarithmic scales  (this is also true for quantities $\Sigma_{\rm t}$ and
$h$), which could be attributed to the fact that $\nu_1$ is indeed
proportional to $\nu_2$ (or one goes from one solution to the other by a
change of $\alpha$; see power law solutions for $\alpha$-discs). To
demonstrate that these two viscosity prescriptions are intrinsically different
(as suggested by what happens in the self-gravitating regions), we have
computed the ratio $\frac{\nu_1}{\nu_2}=\frac{3}{2} \Omega \lambda_p
\Gamma_1/c_{\rm s}$ with $\nu_1$, for the examples discussed above. We have
chosen two radii, the one lies in the classical disc part, the other is in the self-gravitating part. The results are shown in Fig.(\ref{fig:deltanu}).

\begin{figure*}
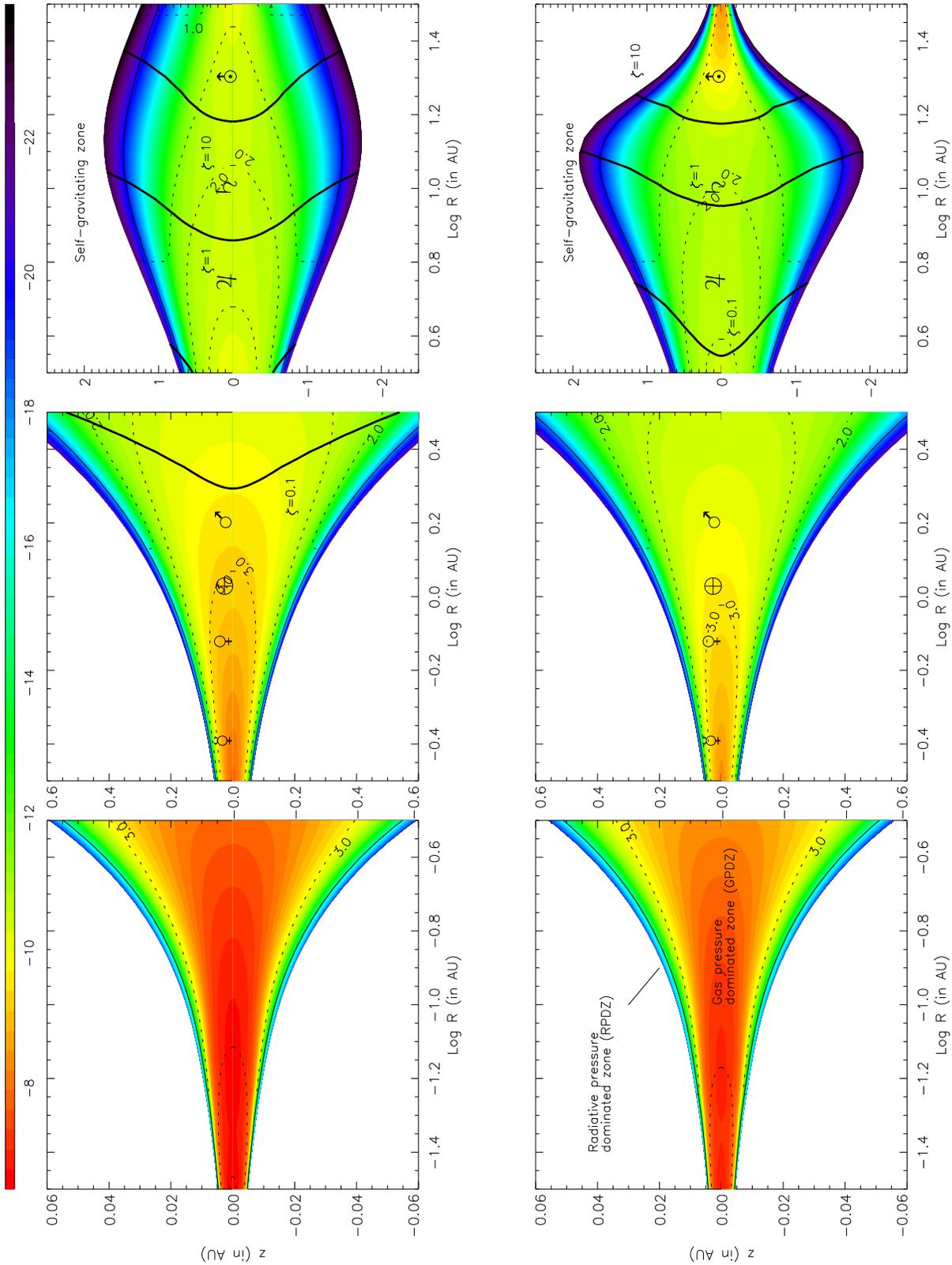

\begin{tabular}{cc}
\psfig{figure=hurejm.f6color,width=7.8cm,angle=90}&
\psfig{figure=hurejm.f6biscolor,width=7.8cm,angle=90}
\end{tabular}
\caption{Density field within a disc surrounding a one solar mass central star, computed with $\nu_1$ ({\it left}) and with $\nu_2$
  ({\it right}). The accretion rate and viscosity parameter are $\dot{M}=10^{-7}$ M$_\odot$/yr and
  $\alpha=10^{-3}$ respectively. The color code refers to $\log \rho$ (with
  $\rho$ in g/cm$^3$) and is the same for both maps. Iso-values of $\log T$
  are given in dotted lines. Also indicated in bold lines are contours for $\zeta=0.1, 1$ and $10$. The limit between the radiative pressure dominated
  zone (RPDZ) and the gas pressure dominated zone (GPDZ) is marked by a
  thin plain line.}
\label{fig:psn}
\end{figure*}

\begin{figure*}
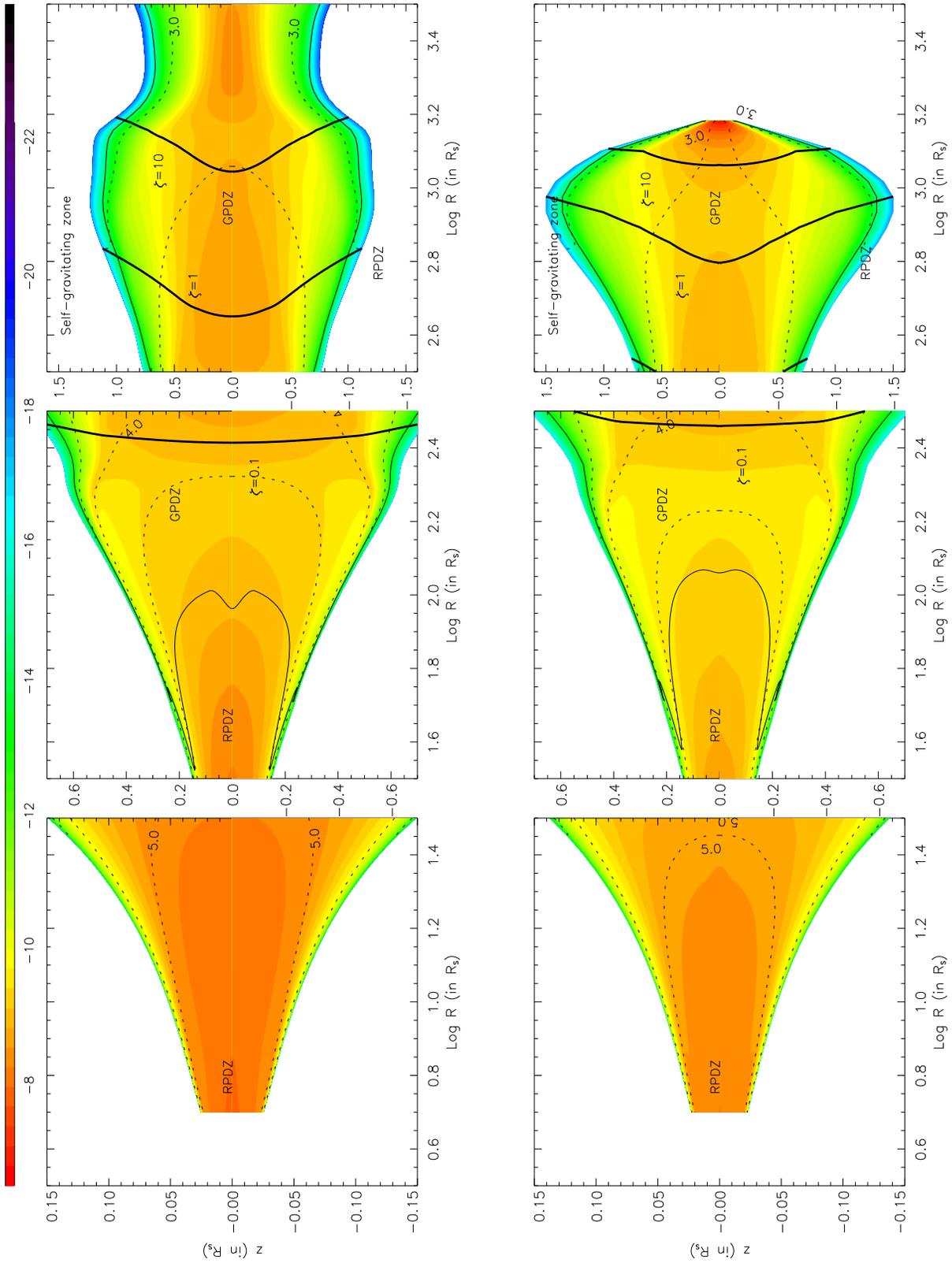

\begin{tabular}{cc}
\psfig{figure=hurejm.f7color,width=7.8cm,angle=90}&
\psfig{figure=hurejm.f7biscolor,width=7.8cm,angle=90}
\end{tabular}
\caption{Same legend (and same color code) as for Fig.(\ref{fig:psn}) but for
  an AGN disc surrounding a $10^8$ M$_\odot$ black hole, with
  $\dot{M}=10^{-2}$ M$_\odot$/yr and $\alpha=10^{-1}$.}
\label{fig:agn}
\end{figure*}

\subsection{Example of internal structure: 2D-density maps}

We show in Fig.(\ref{fig:psn}) the density field within a circumstellar disc
for the same parameter values as in Fig.(\ref{fig:rprofyso}). This simulation has been carried out using $\nu_1$ and $\nu_2$ as well, including
convection, self-gravity and turbulent pressure. We also indicate iso-contours
of the temperature, the limit between gas pressure dominated zones and radiative
pressure dominated ones, and lines where $\zeta(R,z)=\frac{1}{10}, 1$ and
$10$. We notice that the density and the temperature show strong variations in between the midplane to the
photosphere's base and we are far from a vertically isothermal, homogenous
disc as often considered. Note, specially for viscosity law $\nu_1$, the
re-increase of the midplane density in the radial direction (another density
inversion) as self-gravity
gains in importance. It is remarkable that self-gravity does not appear
suddenly (i.e. on a small radial range) but installs gently and steadily over
a very extended domain. For instance, with law $\nu_2$, it contibutes by $10
\%$ in the hydrostatic equilibrium near $2 - 4$ AU and by $50 \%$ at about 7
AU, and all disc quantities are significantly modified over this domain. We
see clearly that the disc density is slightly smaller with $\nu_2$ than
with $\nu_1$ (same color code for both maps).

We notice that the whole
disc is optically very thick, due to its large surface density. It is then likely that
a mean external heating should not change noticeably the temperature and density at the
midplane. From this point of
view, the neglect of irradiation is entirely justified. It is interesting to see that, as long as an
$\alpha$-model can be applied to the Solar Nebula (e.g. Ruden \&
Pollack, 1991; Papaloizou \& Terquem, 1999;
Drouart et al, 1999), our giant planets (displayed on graphs) lie inside the self-gravitating part of
the disc (this is true whatever the accretion rate, but depends on the
$\alpha$-parameter; see the next Section). As already noticed (Ruden \&
Pollack, 1991), this coincidence is somewhat striking and we do not know to
what extent self-gravity could have played a role in the process of the
formation of outer planet and other objects (Lissauer, 1993).

Similar 2D-maps are displayed in Fig.(\ref{fig:agn}) for an AGN disc with
$M=10^8$ M$_\odot$, $\dot{M}=10^{-2}$ M$_\odot$/yr and $\alpha=10^{-1}$, using
the same color table. Note the existence of an inner radiative pressure
dominated region at $R \lesssim 100 \, R_{\rm S}$, common for AGN discs as
well as the great steepness of density gradients near the surface which may
cause numerical difficulties. Also visible is the density
inversion at $R \simeq 160 - 200 \, R_{\rm S}$ which has less amplitude and is
less extended radially in the presence of convection and turbulent pressure.

\subsection{A criterion to check the importance of vertical self-gravity}

We see on both Figs.(\ref{fig:psn}) and (\ref{fig:agn}) that $\zeta(R,\infty)
< \zeta(R,0)$ and iso-values of $\zeta(R,z)$ form very curved lines
(``diabolo''-shape surfaces in 3D). This is important if one wishes to
estimate correctly the position of the self-gravitating regime. For instance,
in the case of the law $\nu_1$ discussed in Fig.(\ref{fig:psn}), $\zeta=1$  in
the midplane at
7 AU (we have $450 \, R_{\rm S}$ in the AGN case depicted in
Fig.(\ref{fig:agn})) whereas this occurs  in the
disc atmosphere at about twice the distance.

\begin{figure}
\psfig{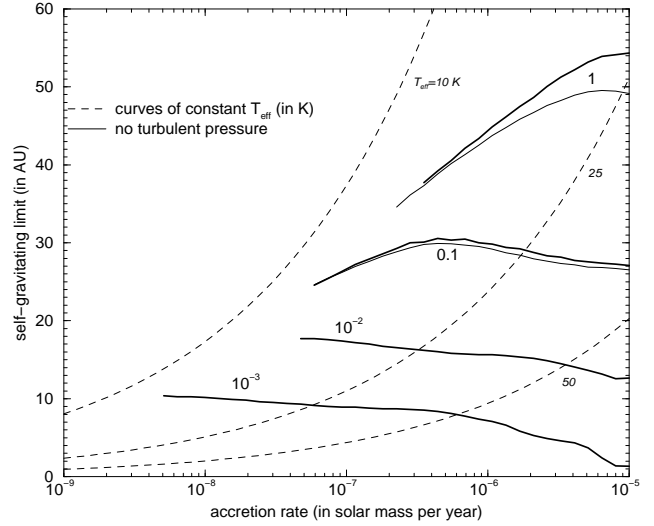}
\caption{Inner edge of the self-gravity dominated disc $R_{\rm sg}$ versus the accretion rate for $\alpha=10^{-3}, 10^{-2}, 0.1$ and $1$ with ({\it bold lines}) and without turbulent pressure ({\it thin lines}). Lines of constant effective temperatures are in dashed lines.}
\label{fig:zetayso}
\end{figure}

For many purposes, it is interesting to know the position of the (vertically)
self-gravitating disc. That is why we have performed a systematic
computation to find the radius $R_{\rm sg}$ satisfying $\zeta(R_{\rm sg},0)=1$ as a function of the accretion rate, for 4 values of the $\alpha$-parameter in the range  $10^{-3}-1$, and for
$M=1$ M$_\odot$ and $M=10^8$ M$_\odot$. These parameter values should cover a
wide variety of circumstellar discs and AGN discs. The calculations have been performed with
$\nu_2$ including convection, self-gravitation, with and without turbulent
pressure (the use of $\nu_1$ would systematically give a
slightly lower value of $R_{\rm sg}$). The results are plotted in Fig.(\ref{fig:zetayso}) for parameter values typical of circumstellar discs. We
see that the location of self-gravitating regime is very sensitive to the
viscosity parameter. The larger the value of $\alpha$, the further away the
self-gravitating region. Regarding the sensitivity to the accretion rate, there
are three different trends. For $\alpha \lesssim 10^{-2}$, the lower
$\dot{M}$, the larger $R_{\rm sg}$. For $\alpha \sim 0.1$, the
dependence is rather weak: $R_{\rm sg} \approx 27 \pm 3$ AU. For
higher values of the viscosity parameter (generally not appropriate to fit
properties of observed discs), $R_{\rm sg}$ is an increasing function of $\dot{M}$. Note also that turbulent pressure, when important, slightly increases $R_{\rm sg}$ ($\sim 12 \%$ in the example for $\alpha=1$ and the highest accretion rate). 

\begin{figure}
\psfig{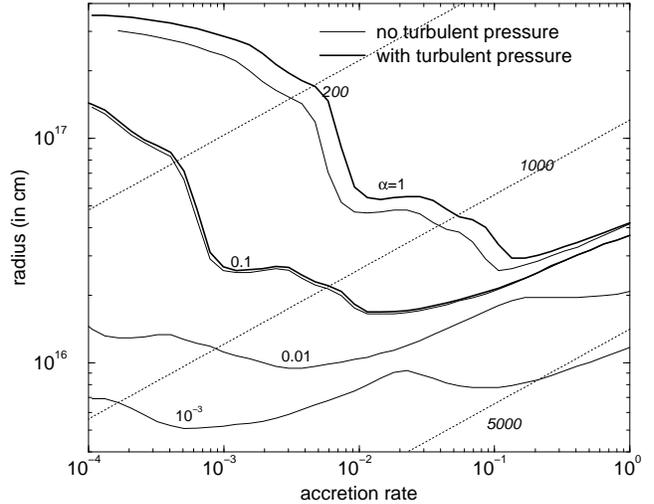}
\caption{Same legend as for Fig.(\ref{fig:zetayso}) but for $M=10^8$ M$_\odot$.}
\label{fig:zetaagn}
\end{figure}

Results for AGN discs are shown in Fig.(\ref{fig:zetaagn}). As above, large
viscosity parameters imply wider non self-gravitating inner disc (because
discs are less dense with large values of the $\alpha$-parameter), and
turbulent pressure pushes further away the self-gravitating regime. For $\alpha
\lesssim 10^{-2}$, we have $R_{\rm sg} \sim 5 - 20 \times 10^{15}$ cm. The
sensitivity to the accretion rate remains weak. For higher values of the
viscosity parameter and for very low to moderate accretion rates ($\dot{M}
\lesssim 10^{-2}-10^{-1}$ M$_\odot$/yr,  depending on $\alpha$), $R_{\rm sg}$ globally decreases as $\dot{M}$ increases. For higher accretion rates, $R_{\rm sg}$ increases as $\dot{M}$ increases and self-gravity is important within the radiative pressure dominated region.

\begin{figure}
\psfig{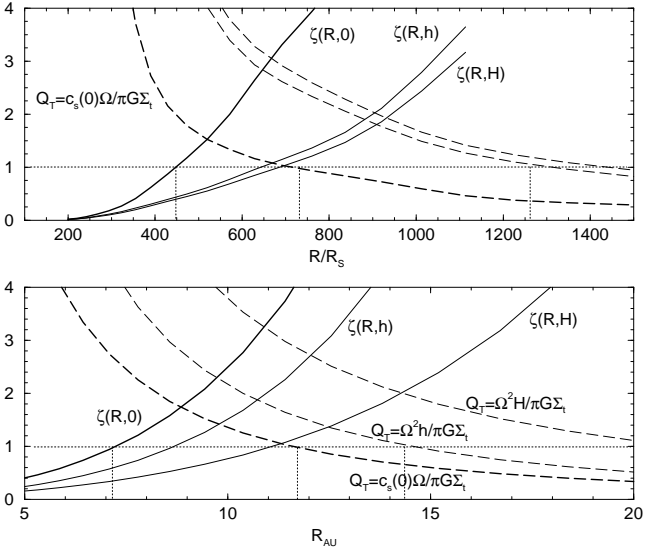}
\caption{Comparison between quantity $\zeta$ at three different altitudes around the
  transition $\zeta=1$ (computed with self-gravity) and the Toomre's parameter
  $Q_{\rm T}$ computed (without self-gravity) for three different
  definitions, for an AGN disc ({\it top}; same parameter values as in
  Fig.(\ref{fig:agn})) and an YSO disc ({\it bottom}; same parameter values as in Fig.(\ref{fig:psn})). }
\label{fig:zetaq}
\end{figure}

It is important to notify that the "Toomre's criterion" ($Q_{\rm T}\equiv
\frac{c_{\rm s}(0) \Omega}{\pi G \Sigma_{\rm t}} \lesssim 1$) is an helpful
tool to trace regions where some gravitational instabilities occur in a
stellar disc (Toomre, 1965), but it differs significantly from the criterion
derived by Goldreich \& Lynden-Bell (1965) for gaseous discs ($Q_{\rm GLB}
\equiv \frac{4 \Omega^2}{\pi G \rho} \lesssim 1$). Besides, $Q_{\rm T}$ is
commonly written in different forms which are not strictly equivalent (e.g., Ruden \& Pollack,
1991; Sincell \& Krolik, 1997; D'alessio, Calvet \& Hartmann, 1997, Papaloizou
\& Terquem, 1999), specially in  a 2D-model. Althought these $Q$-parameters are undisputably related to the $\zeta$-parameter in some ways
(for instance, $Q_{\rm T} \times \zeta(R,\infty) = 2 \frac{c_{\rm
    s}(0)}{\Omega H} > 1$ and $Q_{\rm GLB} \times \zeta(R,0) \simeq 16$), they
must be used with care if one wishes to check the importance of self-gravity. They may lead to noticeable over-estimates of $R_{\rm sg}$, specially if,
as it is always the case, the $Q_{\rm T}$-parameter is computed from models
that discards self-gravity. This is illustrated in
Fig.(\ref{fig:zetaq}) where we have plotted values of $\zeta$ at three key
altitudes and $Q_{\rm T}$ versus the radius. It seems therefore preferable to
make the check on $\zeta(R,0)$ rather than on any other quantities, otherwise
the importance of self-gravity is under-estimated.

\section{Conclusion}

In this article, we have presented the equations accounting for the vertical
structure of steady state keplerian accretion discs, including simultaneously
turbulent pressure, convective transport in the framework of the Mixing Length
Theory, and the disc self-gravity within the infinite slab approximation. The
emphasis has been placed on outer regions where vertical self-gravity becomes
increasingly influencal. We have shown that turbulent pressure makes the disc
thicker, but it can be neglected if $\alpha \lesssim 0.1$. Also, self-gravity may be important even if the disc mass is very low compared to the central
object, contrary to usual assertions. Further, the transition from the
classical disc to the self-gravitating disc takes place gradually and so, concerns a large radial domain.

Another important conclusion of this work is that the model of
mechanical energy deposition towards the vertical direction is
crucial. This is not surprising since viscosity is the source of heating and
accretion. Two different behaviors are predicted inside the
self-gravitating (and gravitationally unstable) region. On the one hand, the
$\alpha \cal{P}$-formalism yields a solution which is
asymptotically cold and diffuse, whereas the local version of the standard
prescription gives a more "singular" solution, asymptotically hot and dense,
which is also predicted by the vertically averaged model. We can imagine that
a suitable choice for the function $\nu(z)$ should provide any intermediate solution, meaning
that this model for outer regions has almost no predictive power. Although
the present stationary keplerian disc model is probably irrelevant to describe
the disc structure in its strongly self-gravitating parts, the existence of two asymptotically distinct
solutions might indicate that the outer disc can get into, either a flat and
cold configuration where fragmentation and formation of indiviual clouds
(Kumar, 1999) and compact objects like planets and stars (Collin \& Zahn,
1999) could occur, or into a thick diffuse configuration (a torus) (Paczynski,
1978). Time dependent simulations should shed light on this question.

This study confirms that discs around super-massive black holes are
self-gravitating close to the center, beyond a few hundreds Schwarzchild
radii, depending on the central mass accretion rate and
$\alpha$-parameter. We have found that the disc surface density remains high
when self-gravity dominates and that the disc mass definitely rises outwards. Let
us remind that a current problem with the fueling of active nuclei is that the accretion time scale $M^{\rm disc}/\dot{M}$ is usually much too long at large radii, that is why other
more efficient mechanisms are invoked (Frank, 1990). However, given the
sensitivity of the disc structure to viscosity, it is not excluded that
depth dependent viscosity laws other than considered here would lead to smaller surface density
distributions, and consequently to shorter accretion time scales. It is
therefore important to
test other models for the energy deposition along the $z$-axis as well as
other kinds of viscosity prescriptions (Lin \& Pringle, 1987; Cannizzo \&
Cameron, 1988; Richard \& Zahn, 1999; see Hur\'e \& Richard, 2000).

With accretion rates and
values of the viscosity parameter usually considered
to model discs in T-Tauri stars, we conclude that our
giant planets (if not Jupiter, the other ones) were probably formed
within the self-gravitating part of the Solar Nebula (Drouart et
al. 1999). This is a fortiori true if these objects experienced any inward migration. More generally, as long as the $\alpha$-theory may be applied to describe the inner parts of circumstellar
discs around forming stars, self-gravity is expected to play a role at
about ten AU from the center. This is not uncompatible with the discs we observed (Beckwith et al., 1990).

\begin{acknowledgements}

I specially thank A. R\'o\.za\'nska for many helpful discussions and for
providing some reference vertical profiles during the writing and test of the
computational code, S. Collin for reading the manuscript. I thank also
D. Gautier for highlights on circumstellar discs and the Primitive Solar
Nebular, J.-P. Zahn for pointing out the role of turbulent pressure. I am
personally grateful to J.M. Hameury for his initiation, some years ago, to the
physics of stellar interiors and vertical structure computations. I also thank
D. Richard for interesting comments on numerical aspects as well as F. Hersant
for widely testing the code. Also, I thank for their hospitality, people at
the ITA-Heidelberg where this work was completed, and specially W.J. Duschl.

\end{acknowledgements}

\appendix
\section{Treatment for convection in the presence of self-gravitation}

Convection is treated with in the framework of the Mixing Length Theory (MLT)
(Cox \& Giuili, 1968). It is modified in order to include turbulent pressure
in the determination of the pressure scale height, and
self-gravititation. Following the original version of the MLT, we neglect the
variation of gravity with the altitude $z$. It is likely that turbulent pressure modify significantly the velocity of rising elements within convectively unstable zones, specially if the $\alpha$-parameter is close to unity, but this effect is not considered here. The total pressure height scale writes
\begin{equation}
\lambda_{\rm p} \equiv - \frac{d z}{d\ln \left( P +p_{\rm t} \right) } \qquad > 0
\label{eq:lambdap}
\end{equation}
where $P$ is the total (gas plus radiation) pressure and $p_{\rm t}$ is the
turbulent pressure. This expression being singular at the midplane ($\lambda_p \rightarrow \infty$) (see Eq.(\ref{eq:eh1})), the pressure scale height is usually limited to the disc thickness
\begin{equation}
\bar{\lambda}_{\rm p} = {\rm Inf} \left[ h, \frac{P +p_{\rm t} }{\rho \left( \Omega^2 z + 4 \, \pi \, G \Sigma \right) } \right]
\label{eq:llambdap}
\end{equation}
where $\rho$ is the gas density, $\Omega$ is the rotation frequency, $\Sigma$
is the surface density and $h$ is the altitude of the bottom atmosphere.

According to Eq.(\ref{eq:temp}), the temperature gradient is
\begin{equation}
\frac{d \ln T}{d z} = - \frac{  \nabla }{\bar{\lambda}_p}
\label{eq:t}
\end{equation}
where the actual gradient $\nabla \equiv \frac{d \ln T}{d \ln P}$ depends on
the adiabatic gradient $\nabla_{\rm ad} = \left( \frac{d \ln T}{d \ln P}
\right)_{\rm ad}$ (computed from the EOS; see the next Section) with respect to the fictitious radiative gradient $\nabla_{\rm r}$ defined as
\begin{equation}
\nabla_{\rm r} = \frac{3 \, \rho \kappa F \bar{\lambda}_p }{16 \, \sigma T^4 }
\label{eq:radgrad}
\end{equation}
where $F$ is the flux to be transported upwards and $\kappa$ is a grey
absorption coefficient which must approach the Rosseland mean $\kappa_{\rm R}$
at great optical depth (see Sect. 3.1).

When $\nabla_{\rm r} \le \nabla_{\rm ad}$ (i.e. in convectively stable zones), heat is transported through radiation only and we have $\nabla = \nabla_{\rm r}$. Conversely, when $\nabla_{\rm r} > \nabla_{\rm ad}$, $\nabla$ is computed from 
\begin{equation}
\nabla = (1 - x^3) \nabla_{\rm r} +x^3 \nabla_{\rm ad}
\end{equation}
where $x$ is the real root of the third degree equation
\begin{equation}
\frac{9}{4} {\cal B}^2 x^3 + {\cal B} x^2 + x -\frac{9}{4} {\cal B}^2 = 0
\end{equation}
with
\begin{equation}
{\cal B} = \left[ \frac{4}{9} {\cal A}^2 \left(  \nabla_{\rm r} - \nabla_{\rm ad} \right) \right]^{1/3}
\end{equation}
and
\begin{equation}
{\cal A} = \frac{1}{48 \sqrt{2} \, \sigma } \frac{c_p \kappa_{\rm R} \alpha^2_{\rm MLT} \lambda^2_p}{T^3} \sqrt{\frac{\rho^5}{P}} \left( \Omega^2 z + 4 \, \pi \, G \Sigma \right)
\end{equation}
where $\alpha_{\rm MLT}$ is the mixing length parameter ($\alpha_{\rm MLT}=1.5$ here) and $c_p$ is the constant pressure specific heat capacity.

\begin{figure}
\psfig{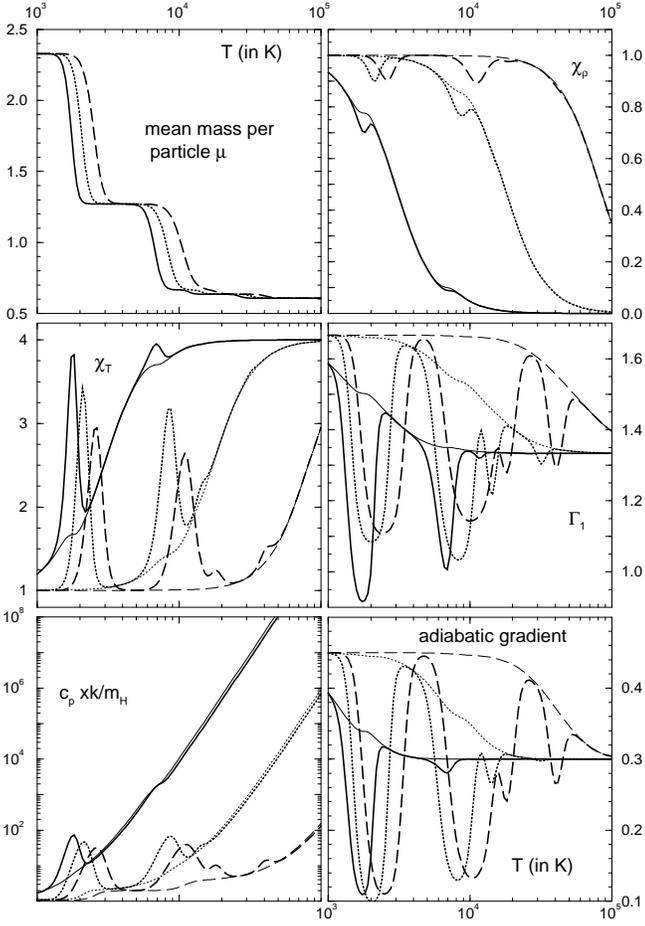}
\caption{Thermodynamic coefficients $\mu$, $\chi_\rho$, $\chi_T$, $c_p$,
  $\Gamma_1$ and $\nabla_{\rm ad}$ computed at thermal equilibrium for a
  hydrogen-helium mixture with elementary abundances H:He=1:$\frac{1}{10}$ for
  $\log \rho=-12$ ({\it plain lines}), $\log \rho=-10$ ({\it dotted
    lines}) and $\log \rho=-8$ ({\it dashed lines}). Raw data are in bold. For $\mu$,
  $\chi_\rho$ and $\chi_T$, the difference between raw data and the high
  precision fitting formula given in the Appendix (Sect. B1) is not visible.
 For all coefficients, thin lines are obtained from low precision approximate expressions (see the Appendix, Sect. B2).}
\label{fig_coefs}
\end{figure}

\section{Notes on the EOS and related quantities}

\subsection{High precision fitting formula for $\mu$ as functions of the temperature and density | $\chi$-coefficients}

\begin{figure}
     \psfig{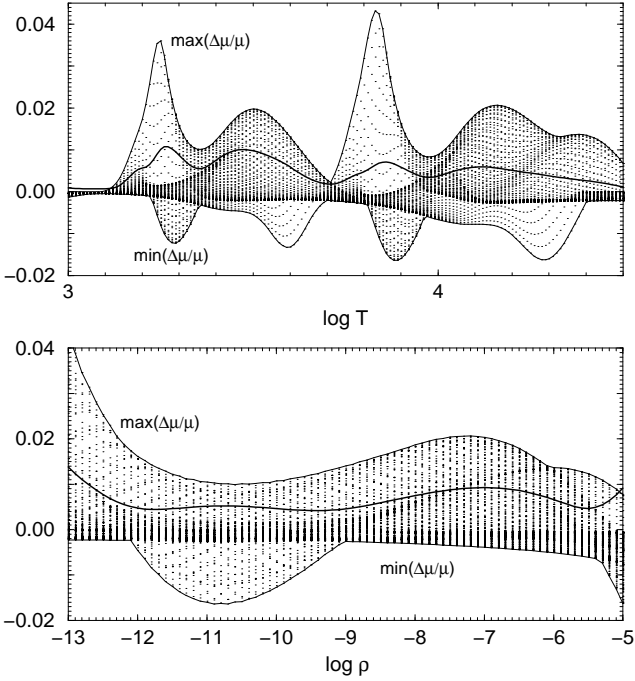}
\caption{Relative error on the mean mass per particle between raw data and fitted values, versus $\log T$ ({\it upper panel}) and versus $\log \rho$ ({\it lower panel}). The standard deviation is plotted in bold line.}
    \label{fig_muerrors}
    \end{figure}

Within the assumption of LTE, the pressure of a mixture of radiation and ideal gas undergoing atomic ionization molecular dissociation is given by Eq.(\ref{eq:eh1}). In particular, the mean mass per particle $\mu$, in units of the proton mass, is defined as
\begin{equation}
\mu  = \frac{ \sum_i {m_i p_i}}{m_{\rm H} \sum_i {p_i}}
\label{eq:mu}
\end{equation}
where $m_i$ and $p_i$ are respectively individual mass and partial pressure of
the chemical compounds, and the sum extends over the total number of
compounds, including electrons. We have computed equilibrium abundances for a
mixture of hydrogen and helium (H:He=1:$\frac{1}{10}$) only as function of
$\rho$ and $T$ (see Hur\'e, 1998). Data relative to the EOS so obtained are
plotted versus the temperature in Fig.(\ref{fig_coefs}) for three values of
the gas density. In particular, the resulting function $\mu(\rho,T)$ has then been fitted with great accuracy by the following expression
\begin{equation}
\mu^{\rm fit} = \delta_0 + \sum_{i=1,4}{\delta_i \tanh \Phi_i},
\label{eq:fitofmu}
\end{equation}
with
\begin{equation}
\Phi_i =  \frac{\log(T) - \theta_i}{ \Delta_i}
\end{equation}
where coefficients $\delta_i$ and $\tau_i$ and $\Delta_i$ are listed in
Tab.(\ref{tab_coefformu}). Hyperbolic functions account successively for the
transitions H$_2$/HI, HI/HII, He/HeI and HeI/HeII. Relative errors on raw data
$\frac{\Delta \mu}{\mu} = \frac{\mu^{\rm fit}-\mu}{\mu}$ are displayed in
Fig.(\ref{fig_muerrors}) and never exceed 4 \% over the whole domain of
temperature and density. Besides, the standard deviations with respect to raw
data are less than 1.3 \%. We have tried to fit the residuals with a few
Gaussians, but no major improvement has been obtained. Note that, by
construction, the fitting formula shows no singular behavior at high of low
densities and temperature. The fit can be used for a cosmic gas without
producing large errors (the effect of heavy elements on the EOS is very weak as long as $Z \ll X$).  

\begin{table}
\caption{Coefficients and functions required to compute $\mu(\rho,T)$ according to Eq.(\ref{eq:fitofmu}). Note that $\delta_1 + ... + \delta_4= 0.618$ (fully ionized helium), and $\delta_0 - (\delta_1 + ... + \delta_4) = 2.373$ (fully molecular gas).}
\begin{tabular}{ll}
i & $\delta_i$  \\ \hline
0 & $+1.4955$ \\
1 & $-0.5400$ \\ 
2 & $-0.3075$ \\
3 & $-0.0160$ \\
4 & $-0.0140$ \\\\
i & $\theta_i$  \\ \hline
1 & $3.93741 + 0.086042 \log \rho + 0.0023141 \log^2 \rho$ \\
2 & $4.74029 + 0.116375 \log \rho + 0.0033417 \log^2 \rho$ \\
3 & $5.07036 + 0.132245 \log \rho +0.0041041 \log^2 \rho $ \\ 
4 & $5.16110 +0.082767 \log \rho + 0.0017907 \log^2 \rho$ \\\\
i & $\Delta_i$ \\ \hline
1 & $0.18303 + 0.020252 \log \rho + 0.0007430 \log^2 \rho$ \\
2 & $0.25730 + 0.030207 \log \rho + 0.0011254 \log^2 \rho$ \\
3 & $0.09435 + 0.006747 \log \rho + 0.0001561 \log^2 \rho$ \\
4 & $0.10794 + 0.009201 \log \rho + 0.0002583 \log^2 \rho $ \\
\end{tabular}
\label{tab_coefformu}
\end{table}

The temperature and density exponents of the total pressure are repectively
given by
\begin{equation}
\chi_T=\left(\frac{\partial {\ln P}}{\partial{ \ln T}}\right)_\rho = 4(1-\beta)+\beta \left[ 1 - \left( \frac{\partial \ln \mu}{\partial \ln T} \right)_\rho \right]
\label{eq:chit}
\end{equation}
and
\begin{equation}
\chi_\rho=\left(\frac{\partial{ \ln P}}{\partial{ \ln \rho}}\right)_T = \beta \left[ 1 - \left( \frac{\partial \ln \mu}{\partial \ln \rho} \right)_T \right]
\label{eq:chirho}
\end{equation}
where $\beta$ is the ratio of gas pressure to total (gas plus radiation) pressure. It follows from Eq.(\ref{eq:fitofmu})
\begin{eqnarray}
\chi_T &=& \frac{\beta}{\mu \log 10} \sum_{i=1,4}{ \frac{\delta_i}{\Delta_i}  \left( 1 -  \tanh^2 \Phi_i \right)}
\nonumber
\\
&+& \beta \left(1 - \delta_0 \right) +4(1-\beta)
\end{eqnarray}
and
\begin{eqnarray}
\chi_\rho &=& \frac{\beta}{\mu \log 10} \sum_{i=1,4}{ \frac{1 -  \tanh^2 \Phi_i}{\delta_i} \left(\frac{d \theta_i}{d \log \rho} \right)_T}
\nonumber
\\
&+& \beta \left(1 - \delta_0  \right)
\end{eqnarray}
and the precision with respect to raw data is also of the order of a few percents.

\subsection{Low precision expressions for coefiicients $\chi_T$, $\chi_\rho$, $\nabla_{\rm ad}$, $\Gamma_1$ and $c_p$}

The adiabatic gradient $\nabla_{\rm ad}$, heat capacity at constant pressure
$c_p$ and adiabatic exponent $\Gamma_1$ are respectively defined by
\begin{equation}
\nabla_{\rm ad} = \left[ \chi_T + \frac{\chi_\rho}{\chi_T} \left( \frac{\partial U}{ \partial \ln \rho} \right)_T \frac{\rho}{P} \right]^{-1}
\end{equation}
where $U$ is the specific internal energy of gas and radiation,
\begin{equation}
c_p = \frac{\chi_T P }{\chi_\rho \rho T \nabla_{\rm ad}}
\end{equation}
and
\begin{equation}
\Gamma_1 =  \frac{\chi_\rho}{1- \chi_T \nabla_{\rm ad}}
\end{equation}

Since disc models are still very uncertrain, one may use, as a first
approximation, low precision expressions for all coefficients related to the
EOS discarding changes in the internal energy due to ionizations and
dissociations, for instance
\begin{equation}
\chi_\rho \sim \beta
\end{equation}
\begin{equation}
\chi_T \sim 4-3\beta
\end{equation}
and then
\begin{equation}
\nabla_{\rm ad} \sim \frac{1}{4-3\beta+\frac{12 \beta}{4-3\beta} \left( 1 -\frac{21}{24}\beta\right)},
\end{equation}
\begin{equation}
c_p \sim \left(\frac{4-3\beta}{\beta^2}\right) \frac{1}{\mu \nabla_{\rm ad}} \frac{k}{m_{\rm H}}
\end{equation}
and
\begin{equation}
\Gamma_1 \sim \frac{\beta}{1-(4-3\beta)\nabla_{\rm ad}}
\end{equation}


\begin{thebibliography}{}
\bibitem{} Abramowicz M.A., Czerny B., Lasota J.P., Szuszkiewicz E., 1988, ApJ, 332, 646
\bibitem{} Abramowicz M.A., Calvani M., Nobili, L., 1980, ApJ, 242, 772
\bibitem{} Adams F.C., Shu F.H., 1986, ApJ, 308, 836
\bibitem{} Aikawa Y., Umebayashi T., Nakano T., Miyam S.M., 1999, ApJ, 519, 705
\bibitem{} Alexander D.R. 1998, private communication
\bibitem{} Alexander D.R., Ferguson J.W., 1994, ApJ, 437, 879
\bibitem{} Artemova I.V., Bisnovatyi-Kogan G.S., Bj\"ornsson G., Novikov I.D. 1996, ApJ, 456, 119
\bibitem{} Bachev, 1999, A\&A,348,71
\bibitem{} Beckwith S.V.W., Sargent A.I., Chini R.S., G\"usten R., 1990, AJ,
  99,924
\bibitem{} Bell, K.R., Cassen P.M., Klahr H.H., Henning Th., 1997, ApJ, 486, 372
\bibitem{} Bell K.R., Lin D.N.C., 1994, ApJ, 427, 987
\bibitem{} Bertin G., Lodato G., 1999, A\&A, 350, 694
\bibitem{} Bertout C., 1989, ARA\&A, 27, 351
\bibitem{} Bodo G., Curir A., 1992, A\&A, 253, 318
\bibitem{} Boss A., 1996, ApJ, 469, 906
\bibitem{} Brandenburg A., 1998, in {\it ``Theory of black hole accretion discs''}, p61, Eds. Abramowicz, Bj\"ornsson \& Pringle J.E., Cambridge University Press
\bibitem{} Burgers P., Lamers H.J.G.L.M, 1989, A\&A, 218, 161
\bibitem{} Burderi L., King A.R., Szuszkiewicz E., 1998, ApJ, 509, 85
\bibitem{} Camenzind M., Demole F., Straumann N., 1986, A\&A, 158, 212
\bibitem{} Cannizzo J.K., 1992, ApJ, 385, 94
\bibitem{} Cannizzo J.K., 1993, in {\it ``Accretion disks ionm compact stellar systems''}, p6, Ed. J.C. Weehler, World Scientific
\bibitem{} Cannizzo J.K., Reiff C.M. 1992, ApJ, 385, 87
\bibitem{} Cannizzo J.K., Cameron A.G.W., 1988, 330, 327
\bibitem{} Cannizzo K.K., Wheeler J.C., 1984, ApJ Sup. Ser., 55, 367
\bibitem{} Collin-Souffrin S., 1987, A\&A, 179, 60
\bibitem{} Collin-Souffrin S., Dumont A.-M. 1990, A\&A, 229, 292
\bibitem{} Collin S., Zahn J.P., 1999, A\&A, 344, 433
\bibitem{} Collin S., Hur\'e J.M., 1999, A\&A, 341, 385
\bibitem{} Cox J.P., Giuli R.T., 1968, in {\it ``Principles of stellar structure''}, New York, Gordon and Breach
\bibitem{} Chiang E.I., Goldreich P., 1997, ApJ, 490, 368
\bibitem{} Chick K.M., Cassen P., 1997, ApJ, 477, 398
\bibitem{} Clarke C.J. 1988, MNRAS, 235, 881
\bibitem{} D'Alessio P., Calvet N., Hartmann, 1997, ApJ, 474, 397
\bibitem{} D'Alessio P., Canto G., Calvet N., Lizano S., 1998, ApJ, 500, 411
\bibitem{} D'Alessio P., Cant\'o J., Hartmann L., Calvet N., Lizano S., 1999, ApJ, 511, 896
\bibitem{} D\"orrer T., Riffert H., Staubert R, Ruder H., A\&A 1996, 311, 69
\bibitem{} De Kool M., Wickramasinghe D., 1999, MNRAS, 307, 449
\bibitem{} Drinkwater M.J., Combes F., Wiklind T., 1996, A\&A, 312, 771
\bibitem{} Drouart A., Dubrulle, B., Gautier, D., Robert F., 1999, I, 140, 129
\bibitem{} Dubus G., Lasota J.P., Hameury J.M., Charles P., 1999, MNRAS, 303,139
\bibitem{} Duley W.W., Williams D.A., in {\it ``Interstellar chemistry''}, London, England and Orlando, FL, Academic Press, 1984
\bibitem{} Duschl W.J, Strittmatter P.A., Biermann P.L., 2000, in press
\bibitem{} Duvert G. et al., 1998, A\&A, 332, 867
\bibitem{} El-Khoury Walid, Wickramasinghe D., 1999, MNRAS, 303, 380
\bibitem{} Eriguchi Y., M\"uller, 1991, 248, 435
\bibitem{} Falcke H., 1998, Rev. Mod. Astr., 11, Schielicke R.E. ed.
\bibitem{} Frank A., 1998, in {\it''Accretion processes in astrophysical systems: some like it hot''}, Proceedings of the 8th AIP Conference, Ed. S.T. Holt, Kallman T.R., 513
\bibitem{} Frank J., King A., Raine D. 1992, in {\it ``Accretion power in
    astrophysics''}, 2nd Ed., Cambridge Uni. Press.
\bibitem{} Fukue J., 1992, PASJ, 44, 663
\bibitem{} Goldreich P., Linden-Bell D. 1965, MNRAS, 130, 97 
\bibitem{} Guilloteau S., Dutrey A., 1998, A\&A, 339, 467
\bibitem{} G\"usten R., Chini R., Neckel T., 1984, 138, 205
\bibitem{} Hameury J.M. et al., 1998, MNRAS, 298, 1048
\bibitem{} Hashimoto M., Eriguchi Y., M\"uller E., 1995, A\&A, 297, 135
\bibitem{} Henning \& Stognienko R., 1996, A\&A 311, 291
\bibitem{} Herrnstein J.R. et al., 1999, Nature, 400, 539
\bibitem{} Hersant F., Gautier D., Hur\'e J.M., 2000, in preparation
\bibitem{} Hubeny I. 1990 ApJ, 351, 632
\bibitem{} Hubeny I., Hubeny V., 1998, ApJ, 505, 558
\bibitem{} Hunter C., 1963, MNRAS, 126, 23
\bibitem{} Hur\'e J.M., Collin-Souffrin S., Le Bourlot J., Pineau des For\^ets G. 1994a, A\&A, 290, 19
\bibitem{} Hur\'e J.M., Collin-Souffrin S., Le Bourlot J., Pineau des For\^ets G. 1994b, A\&A, 290, 34                                                         \bibitem{} Hur\'e J.M. 1997,  {\it''Accretion processes in astrophysical systems: some like it hot''}, Proceedings of the 8th AIP Conference, Ed. S.T. Holt, Kallman T.R., 137
\bibitem{} Hur\'e J.M. 1998, A\&A, 337, 625
\bibitem{} Hur\'e J.M., Richard D., to appear in {\it "AGN in their cosmic
    environment"}, Eds. B. Rocca-Volmerange, Sol H., EDPS Conference Series in
  Astronomy \& Astrophysics
\bibitem{} Hur\'e J.M., Richard, Zahn J.P., 2000, in preparation
\bibitem{} Igea J., Glassgold E, 1999, ApJ, 518, 848
\bibitem{} Kenyon S.J., Hartmann L., 1987, ApJ, 323, 714
\bibitem{} Kenyon S.J., Yi I., Hartmann L., 1996, ApJ, 462, 439
\bibitem{} Kippenhahn R., Weigert A., 1990, in {\it "Stellar structure and
    evolution"}, Springer Berlin
\bibitem{} Kumar P., 1999, ApJ, 519, 599
\bibitem{} Lasota J.-P., Hameury J.-P. 1998, in {\it''Accretion processes in astrophysical systems: some like it hot''}, Proceedings of the 8th AIP Conference, Ed. S.T. Holt, Kallman T.R., 351
\bibitem{} Laughlin G., R\'o\.zyczka M., 1996, ApJ, 456, 279
\bibitem{} Lin D.N.C., Papaloizou J.C.B., 1980, MNRAS, 191, 37
\bibitem{} Lin D.N.C., Pringle J.E., 1987, MNRAS, 225, 607
\bibitem{} Liu B.F., Xie G.Z., Ji K.F., 1994, A\&SS, 220, 75
\bibitem{} Lynden-Bell D., Pringle J. 1974 MNRAS, 168, 603
\bibitem{} Lissauer J.J., 1993, ARAA, 31, 129
\bibitem{} Malbet F., Bertout C., 1991, 383, 814
\bibitem{} Maraschi L., Reina C., Treves A., 1976, ApJ, 206, 295
\bibitem{} Masuda N., Nishida S. Eriguchi Y., 1998, MNRAS, 297, 1139
\bibitem{} Meyer F., Meyer-Hodmeister E., 1982, A\&A, 106, 34
\bibitem{} Miyoshi M. et al. 1995, Nature, 373, 127
\bibitem{} Mihalas D., 1978, in {\it ``Stellar Atmospheres''} (San Fransisco: Freeman)
\bibitem{} Milsom J. A., Chen X., Taam R. E., 1994, ApJ, 421, 668 
\bibitem{} Mineshige S., Osaki Y., 1983, PASJ, 35, 377
\bibitem{} Mineshige S., Tuchman Y., Wheeler J.C., 1990, ApJ, 359, 176
\bibitem{} Mineshige S., Umemura. M, 1997, ApJ, 480, 167
\bibitem{} Mundy L.G., Looney L.W. and Welch, W.J., 2000, in {\it ``Protostars and Planets IV''}, Ed. Mannings V., Boss A.P., Russell S.S. (Tucson: University of Arizona Press), in press 
\bibitem{} Narayan R., Madevan R., Quataert E., 1998, in {\it ``The Theory of Black Hole Accretion Disks''}, Eds. Abramowicz M.A., Bjornsson G. \& pringle J.E., Cambridge Uni. Press
\bibitem{} Nayakshin S., Kazanas D., Kallman, T., 1999, AAS, 195, 3903
\bibitem{} Osterbrock D.E., 1993, Ap.J., 404, 551
\bibitem{} Paczy\'nski B. 1978, AcA, 28, 91
\bibitem{} Papaloizou J.C.B., Terquem C., 1999, ApJ, 521, 823
\bibitem{} Pojma\'nski G., 1986, Acta A, 36, 69
\bibitem{} Pollack J.B. et al. 1994, Ap. J., 421, 615
\bibitem{} Press W.H., Teukolsky S.A., Vetterling W.T., Flannery B.P., 1992, in {\it ``Numerical recipes in FORTRAN. The art of scientific computing''}, Cambridge: University Press, p727
\bibitem{} Pringle J. 1981 ARA\&A, 19, 137
\bibitem{} Pringle J.E., Rees  M.J. 1972, Astr.  Ap., 21, 1
\bibitem{} Pudritz R.E. et al., 1996, ApJ, 470, L123
\bibitem{} Richard D., Zahn J.-P. 1999, A\&A, 347, 734
\bibitem{} Ross R.R., Fabian A.C., Mineshige S., 1992, MNRAS, 258, 189
\bibitem{} Robinson E.L., MArsh T.R., Smak J.I., 1993, in {\it ``Accretion disks ionm compact stellar systems''}, p75, Ed. J.C. Weehler, World Scientific
\bibitem{} R\'oza\'nska A., Czerny B., Zycki P.T., Pojma\'nski G., 1999, MNRAS, 305, 481
\bibitem{} R\'oza\'nska A. 1998, private communication
\bibitem{} Ruden S.P., Pollack J.B., 1991, ApJ, 375, 740
\bibitem{} Sakimoto P.J., Coroniti F.V. 1981, ApJ, 247, 19
\bibitem{} Sanders D.B. et al. 1989, ApJ, 347, 29 
\bibitem{} Sandqvist Aa., 1999, A\&A, 343, 367
\bibitem{} Seaton M.J., Yan Y., Mihalas D., Pradhan A.K., 1994, MNRAS, 266, 805
\bibitem{} Shakura N.I., Sunyaev R.A. 1973, A\&A, 24, 337
\bibitem{} Shaviv G., Wehrse R., 1991, A\&A, 251, 117
\bibitem{} Shepherd D.S., Kurtz S.E., 1999, ApJ, 523, 690
\bibitem{} Shields G.A., 1977, Ap.J. Let., 18, 119
\bibitem{} Shlosman I., Begelman M.C. 1987, Nature, 329, 810
\bibitem{} Shore S.N., White R.L. 1982 ApJ. 256, 390
\bibitem{} Shu F. H., Tremaine S., Adams F. C., Ruden S. P., 1990, ApJ, 358, 495
\bibitem{} Siemiginowska A., Czerny B., Kostyunin V. 1996, ApJ, 458, 491
\bibitem{} Sincell M., Krolik J., 1997, ApJ, 476, 605
\bibitem{} Smak J., 1984, Acta Astron., 34, 161
\bibitem{} Stahler S.W., 1983, ApJ, 268, 155
\bibitem{} St\"orzer H., 1993, ApJ, 271, 25
\bibitem{} Terquem C., Bertout C., 1993, A\&A, 274, 291
\bibitem{} Toomre A. 1964, ApJ, 139, 1217
\bibitem{} Th\'e P.S., Molster F.J., 1994, A\&SS, 212, 125
\bibitem{} Tuchman Y., Mineshige S., Wheeler J.C., 1990, ApJ, 359, 164 
\bibitem{} Voit G.M., 1991, ApJ, 379, 122
\bibitem{} Wehrse R., St\"orzer H., Shaviv G.,, 1993, A\&SS, 205, 163
\bibitem{} Zeippen C.J., 1998, private comminucation
\bibitem{} Zdziarski A.A. 1986, ApJ, 305, 45
\end{thebibliography}
\end{document}